\documentclass[journal, twoside]{IEEEtran}

\usepackage{comment}

\usepackage[dvipsnames]{xcolor}
\usepackage{enumerate}
\usepackage{amssymb}
\usepackage{amsmath}
\usepackage{soul}
\usepackage{epsfig}
\usepackage{multirow, makecell}
\usepackage{verbatim}
\usepackage{epsfig}
\usepackage{amsmath, amssymb}
\usepackage{subfigure}
\usepackage{times}
\usepackage{pifont}
\usepackage{adjustbox}
\usepackage{array,multirow,graphicx}
\usepackage[ruled,linesnumbered]{algorithm2e}
\SetAlFnt{\small}
\usepackage{longtable}
\usepackage{textcomp}
\usepackage{rotating}
\DontPrintSemicolon

\usepackage{chemformula}
\usepackage{xcolor}
\usepackage{pgfplots}
\usepackage{multicol}
\usepackage{mathrsfs}
\usepackage{color}
\usepackage{tikz}
\usetikzlibrary{plotmarks,shapes,decorations}
\usepackage{pgfplots}
\usepackage{listings}
\usepackage{xspace}
\usepackage{newtxmath}

\usepackage{enumitem} 
\usepackage{tikz}
\newcommand*\circled[1]{\tikz[baseline=(char.base)]{
            \node[shape=circle,draw,inner sep=1pt] (char) {#1};}}
            
\usepackage{xurl}
\usepackage[switch]{lineno}

\usepackage{hyperref}
\hypersetup{
    colorlinks=true,
    linkcolor=blue,
    filecolor=blue,
    urlcolor=black,
    citecolor=blue
}

\usepackage[hyphenbreaks]{breakurl}

\newcommand{\ie}{{\em i.e.,}\xspace}

\begin{document}
\bstctlcite{bstctl:nodash}
\title{Toward Data-Driven Digital Therapeutics Analytics: Literature Review and Research Directions}

\author{Uichin Lee, Gyuwon Jung, Eun-Yeol Ma, Jin San Kim, Heepyung Kim,\\Jumabek Alikhanov, Youngtae Noh, Heeyoung Kim
\thanks{U. Lee, G. Jung, E. Ma, H. Kim, and H. Kim are affiliated with KAIST. J. S. Kim and J. Alikhanov are affiliated with Inha University. Y. Noh is affiliated with KENTECH.}%
\thanks{We thank Dr. Yoonsup Choi for carefully reviewing this manuscript and providing constructive feedback.}}

\markboth{IEEE/CAA JOURNAL OF AUTOMATICA SINICA,~Vol.~X, No.~X, X~X}{Lee \MakeLowercase{\textit{et al.}}: Toward Data-Driven Digital Therapeutics Analytics: Literature Review and Research Directions}

\maketitle

\begin{abstract}  
With the advent of Digital Therapeutics (DTx), the development of software as a medical device (SaMD) for mobile and wearable devices has gained significant attention in recent years. Existing DTx evaluations, such as randomized clinical trials, mostly focus on verifying the effectiveness of DTx products. To acquire a deeper understanding of DTx engagement and behavioral adherence, beyond efficacy, a large amount of contextual and interaction data from mobile and wearable devices during field deployment would be required for analysis. In this work, the overall flow of the data-driven DTx analytics is reviewed to help researchers and practitioners to explore DTx datasets, to investigate contextual patterns associated with DTx usage, and to establish the (causal) relationship between DTx engagement and behavioral adherence. This review of the key components of data-driven analytics provides novel research directions in the analysis of mobile sensor and interaction datasets, which helps to iteratively improve the receptivity of existing DTx.
\end{abstract}

 
\IEEEpeerreviewmaketitle

\section{Introduction}
Digital therapeutics (DTx), unlike traditional treatments such as pills, uses software installed in smartphones or wearable devices as software as a medical device (SaMD) to cure diseases and improve health conditions, which is a major departure from existing wellness products (e.g., Fitbits)~\cite{sverdlov2018digital}. As with traditional therapeutics, DTx also requires clinical validation of efficacy through systematic clinical trials~\cite{dta2018dtx}. 

The US FDA has already authorized a number of digital therapeutic products, for example, WellDoc's BlueStar~\cite{Quinn:2011} for diabetes management, and Pear Therapeutics' reSET~\cite{Campbell:2014} for drug addiction recovery, opening up new DTx possibilities, such as doctors' prescriptions and insurance reimbursement. Unlike the traditional drug development, the cost of DTx development is relatively low, and new DTx markets are growing rapidly. The DTx Alliance, which was formed in 2017, consists of both startups (e.g., Omada Health~\cite{Sepah:2017} and Akili~\cite{kollins2020novel}) and global pharma (e.g., Novartis and Bayer). The DTx market is estimated to increase to \$8.7 billion in 2025, with an average annual growth rate of 20\%~\cite{GrandView2017}.

DTx therapies mostly consider behavioral changes in chronic diseases (e.g., diabetes and cardiovascular diseases)~\cite{greenwood2017systematic, widmer2015digital} and neuropsychiatric diseases (e.g., depression, sleep disorders, and attention deficit hyperactivity disorder (ADHD))~\cite{hollis2017annual}. These are the areas in which the treatment effects of cognitive behavior therapies are significant. DTx therapies can deliver patient-centered care by supplementing the areas in which treatment is difficult or poorly managed through existing treatment methods (e.g., lifestyle coaching and cognitive behavior therapies), to improve the quality of care at lower costs.

One of the important components of traditional drug development is the selection and optimization of a drug delivery system that aims to effectively deliver a specific drug to the desired target (e.g., sustained release with microneedle patches)~\cite{Liu2016DDS}. Digital therapeutics can deliver various interventions through digital technologies (e.g., interactive mobile content, videos, chatbots, and push notifications)~\cite{mohr2014behavioral, lee2019intelligent}. Thus, it is very important to analyze and optimize the engagement and receptivity of ``DTx delivery systems'' using mobile and wearable devices.

The existing drug delivery systems can be evaluated in controlled environments. However, DTx usage is a daily occurrence in the lives of patients, and thus, it is very difficult to evaluate the real-world user experiences and the efficacy of DTx in a laboratory setting~\cite{lee2019intelligent}. Traditional clinical trials on DTx mostly focus on measuring the endpoints or proximal/distal outcomes in the wild, but less attention has been paid to systematically understanding DTx user engagement and adherence patterns, which are essential for DTx improvement. Furthermore, there is a lack of agreement on the methods and criteria for evaluating DTx related user experiences~\cite{Wiederhold2021}.

\begin{figure*}[t]
\centering
\includegraphics[width=\textwidth]{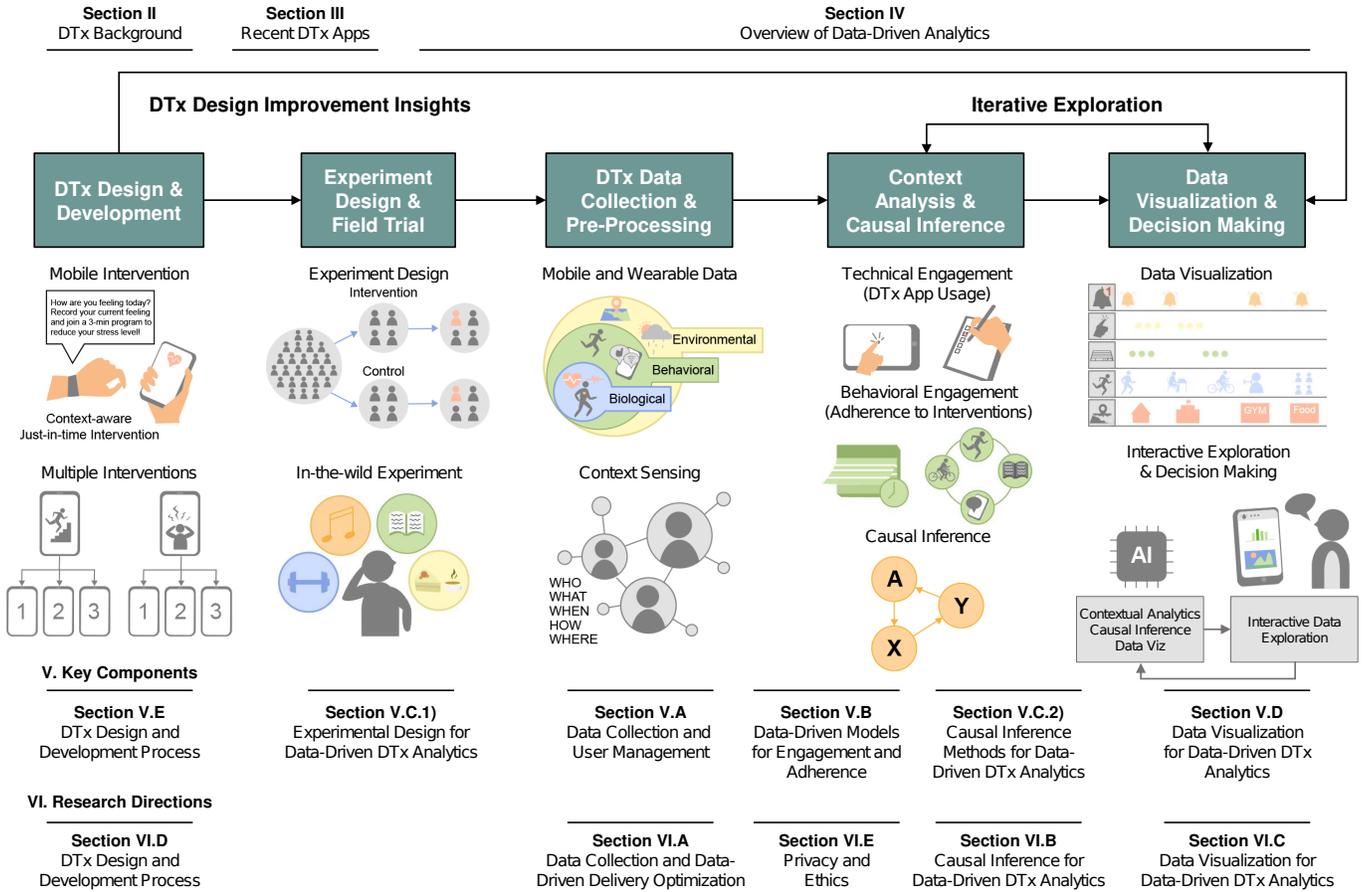}
\caption{A conceptual diagram of data-driven DTx analytics: a process overview (Section IV) and key component reviews (Section V)} 
\label{fig:dtxprocess}
\end{figure*}

The receptivity to DTx relates to the overall process of intervention delivery using digital devices (e.g., notification delivery, notification perception/checking, and behavioral adherence)~\cite{choi2019multi}. DTx aims to induce behavioral changes in users, and it is very important to analyze patient DTx engagement and receptivity to shorten the DTx development time and to maximize the effectiveness of DTx. This review aims to illustrate the flow and key components of data-driven DTx analytics to help researchers and practitioners to investigate the user engagement and intervention receptivity to DTx by analyzing digital footprint data (known as digital phenotype data) collected from mobile and wearable devices. Data-driven DTx analytics for user engagement and intervention receptivity in DTx delivery systems will provide key insights for the improvements of DTx, possibly innovating the existing paradigm of DTx development processes.

Data-driven DTx analytics is closely related to automation research because DTx replaces human-based health interventions with mobile-based counterparts. Furthermore, intelligent agents leverage context sensing and data processing to automate and personalize health services. This topic is also related to human-in-the-loop system design for proactive guiding of user behavior where an operator's behavior and machine intelligence act as human-cyber-physical systems that are critical for automation performance~\cite{ji2019}. Automation fields have a long tradition of using sensor data to optimize the performance of human-machine collaboration. This work on data-driven DTx analytics broadens the scope of existing automation research.

The remainder of this paper is organized as follows (Fig. \ref{fig:dtxprocess}). In Section II, background information (e.g., DTx definition and regulations) is provided. In Section III, an overview of recent DTx is provided, and opportunities for adopting just-in-time interventions in DTx are discussed. In Sections IV and V, an overview of the flow of data-driven DTx analytics and a detailed review of key components are provided. In Section VI, to conclude, several directions for future research are discussed.

\section{DTx Background}
\subsection{Defining DTx and its Relationship to Digital Health}

The Digital Therapeutics Alliance, an industry association in the digital health area, defines DTx as ``evidence-based therapeutic interventions to patients that are driven by high-quality software programs to prevent, manage, or treat a medical disorder or disease''~\cite{dta2018dtx}. In academia, DTx is similarly defined: e.g., ``a new treatment modality in which digital systems (e.g., smartphone apps) are used as regulatory body-approved, prescribed therapeutic interventions to treat medical conditions''~\cite{sverdlov2018digital}. Existing DTx mostly target chronic diseases with continuous intervention support to change behaviors or lifestyles and to help patients manage their health conditions effectively. The coverage of therapeutics includes obesity, prediabetes, hypertension, hyperlipidemia, smoking-related diseases, chronic pain, chronic obstructive pulmonary disease and asthma, diabetes, alcoholism, coronary artery disease, and serious mental illnesses~\cite{kvedar2016digital, Patel2020}.

Although DTx has recently been in the spotlight, the use of digital products in providing therapeutic interventions has been studied for a long time. As Webb et al.~\cite{webb2010using} state in their study, researchers have designed therapeutic interventions on theoretical bases, applied behavioral change techniques in them, and delivered them via the Internet. The term ``digital therapeutics'' was first mentioned in a study published in 2015~\cite{sepah2015long}, defining it as ``evidence-based behavioral treatments delivered online that can increase accessibility and effectiveness of health care.'' As this definition implies, it became important to verify the effectiveness of DTx treatments (i.e., the actual outcome of the DTx treatment in the real world)~\cite{capobianco2015digital, sepah2017double}. Additionally, the industry has also begun to set up standards for clinical evidence.

Digital therapeutics is deeply related to digital health and mHealth and is generally classified as a subset of digital health. According to the World Health Organization, the origin of digital health is eHealth, which is defined as ``the use of information and communications technology in support of health and health-related fields.'' This definition includes mHealth, which is ``the use of mobile wireless technologies for health.'' As a subset of eHealth, mHealth expands its scope to emerging technologies, such as big data analysis, artificial intelligence, and genomics~\cite{world2019guideline}. The US FDA, on the other hand, extends the area of digital health to cover ``mobile health (mHealth), health information technology, wearable devices, telehealth, telemedicine, and personalized medicine''~\cite{fda2020digitalhealth}. There is another subset of digital health named ``digital medicine,'' which is defined as ``high-quality hardware and software that support the practice of medicine broadly'' for measurement and intervention in health-related services~\cite{dime2021, dta2018industry}. For instance, digital medicine may include pills with built-in sensors for monitoring tumors inside the body or wearables that continuously track the glucose level or heart rate. Healthcare providers or consumers then utilize the collected data to manage the health status by changing treatment decisions or medication doses. In this sense, digital medicine can be seen as a broader concept than DTx~\cite{aungst2021}. Compared with digital medicine, most DTx products focus only on software and emphasize making direct changes in health conditions or adjusting treatments based on the collected health data.

Digital health uses technology to deliver information and enable communication, with the purpose of monitoring and managing patients and consequently improving their health conditions~\cite{iyawa2016digital}. Moreover, this approach reduces the burden of health care, allows patients to manage their health even outside traditional hospitals, and individualizes the treatment by implementing behavior change theory or techniques~\cite{widmer2015digital}. Following these definitions and descriptions, digital therapeutics can be considered part of digital health. In particular, during the COVID-19 era, the US government relaxed the regulations for noninvasive remote monitoring devices and reduced direct contact between patients and healthcare providers~\cite{FDA-NonInvasive}. This is expected to increase the use of DTx in everyday life and eventually address health inequalities by supporting patients who cannot easily access medical facilities, such as people living in rural areas~\cite{dta2020rural}.

Various studies on DTx have been conducted in the area of digital health using different digital platforms (e.g., web and mobile apps). Digital health research targets many diseases or health issues, such as diabetes~\cite{greenwood2017systematic}, cardiovascular diseases~\cite{widmer2015digital}, asthma~\cite{morrison2014digital}, mental health (e.g., depression, anxiety, ADHD, autism spectrum disorders, and eating disorder)~\cite{hollis2017annual}, weight management~\cite{khaylis2010review}, and smoking cessation~\cite{ghorai2014mhealth}. In addition, digital health solutions could be utilized to manage various types of cancer tumors~\cite{aapro2020digital}, support patients in rehabilitation from neurological diseases (e.g., stroke, Parkinson's disease, and multiple sclerosis)~\cite{choi2019digital, abbadessa2021digital}, and prevent and treat HIV~\cite{simoni2015opportunities, PositiveLinks}. A report published in 2017 showed that there are more than 318,000 digital health apps publicly available to consumers through Apple Store and Google Play, with more than 200 apps added each day~\cite{aitken2017growing}.

However, not all apps can be classified as ``digital therapeutics'' because clinical evidence and real-world outcomes (e.g., managing, preventing, or treating a medical disorder, or optimizing medication) have to be satisfied for regulatory purposes. Therapeutic interventions must also be certified by regulatory bodies in terms of efficacy and safety for intended use~\cite{dta2018industry}; by demonstrating clinical evidence (or efficacy) via randomized controlled trials (RCTs) in which participants are randomly assigned to the clinical interventions or the control group set up for comparison; the control could be a placebo or no intervention at all~\cite{aitken2017growing}. Research communities on digital health have been striving to conduct RCTs to obtain clinical evidence. However, the generation of clinical evidence for efficacy and safety is still scarce because of the time and cost of running large-scale RCTs, which has become one of the main challenges in DTx. Further issues might arise such as the technology becoming outdated during the trial, ethical considerations, infeasible placebo control groups, or privacy concerns arising from remote informed consent~\cite{sverdlov2018digital}.

\begin{figure}[!t]
\centering
\includegraphics[width=\columnwidth]{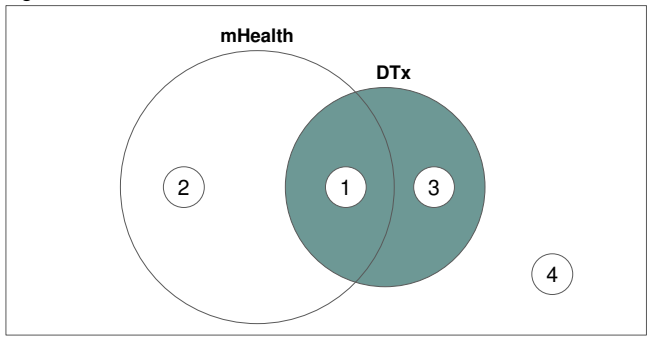}
\caption{Relation among Digital Health, mHealth, and Digital Therapeutics (DTx). There are four different categories in Digital Health depending on whether the system utilizes mobile devices (mHealth) and whether it has sufficient direct therapeutic intervention or clinical evidence (DTx).}
\label{fig:dtxdiagram}
\end{figure}

To summarize, the relationship between digital health, mHealth, and DTx can be described as shown in Fig. \ref{fig:dtxdiagram}. Both mHealth and DTx are subsets of digital health and overlap in some areas.
\begin{enumerate}[label=\protect\circled{\arabic*}]
\item mHealth and DTx: DTx products that deliver therapeutic interventions via mobile devices are included in this area. Most mobile applications and major DTx products use this format.
\item mHealth but not DTx: Digital health apps are available on mobile devices (usually smartphones), but they do not demonstrate clinical evidence or real-world outcomes.
\item DTx but not mHealth: DTx products that do not utilize mobile or wearable platforms. These products may be in the form of software applications or web applications.
\item Digital health, neither mHealth nor DTx: Digital health apps available as a form of software application or web application (not smartphones), but have only the ability to capture, store, display, or transmit health data and information with no direct therapeutic interventions, clinical evidence, or real-world outcomes (e.g., health information technology, telehealth, and medical imaging)
\end{enumerate} 
 
\begin{table*}[]
\centering
\caption{Recent Digital Therapeutics}
\label{dtx}
\scalebox{1}{
\begin{tabular}{llllll}
\hline
\textbf{Category} & \textbf{Disorder/Disease} & \textbf{DTx name} & \textbf{Company} & \textbf{Status} & \textbf{Intention} \\ \hline\hline
\multirow{7}{*}{\begin{tabular}[c]{@{}c@{}}Mental\\ disorder\end{tabular}} & ADHD & EndeavorRx~\cite{kollins2020novel} & Akili Interactive Labs & \begin{tabular}[c]{@{}c@{}}FDA De Novo\end{tabular} & Treat \\ \cline{2-6} 
 & \multirow{2}{*}{Depression} & deprexis~\cite{Meyer:2009} & GAIA-AG & FDA 510 (K) & Manage \\ \cline{3-6} 
 &  & Woebot~\cite{Fitzpatrick:2017} & Woebot Health & \begin{tabular}[c]{@{}c@{}}-\end{tabular} & Manage \\ \cline{2-6} 
 & Insomnia & Somryst~\cite{Ritterband:2009} & Pear Therapeutics & FDA 510 (K) & Manage \\ \cline{2-6} 
  &  \multirow{2}{*}{Panic disorder} & FREESPIRA~\cite{Tolin:2017} & Freespira & FDA 510 (K) & Manage \\ \cline{3-6}  &  & NightWare~\cite{NightWare:2021} & NightWare & FDA De Novo & Manage \\  \cline{2-6}
 & \multirow{2}{*}{Substance abuse} & reSET~\cite{Campbell:2014} & Pear Therapeutics & FDA De Novo & Treat \\ \cline{3-6} 
 &  & reSET-O~\cite{Christensen:2014} & Pear Therapeutics & FDA De Novo & Treat \\ \hline
\multirow{14}{*}{\begin{tabular}[c]{@{}c@{}}Physical\\ disorder\end{tabular}} & \multirow{2}{*}{Asthma} & Propeller~\cite{Merchant:2016} & Propeller Health & FDA 510 (K) & Optimize \\ \cline{3-6}  &  & NuvoAir~\cite{NuvoAir:2021} & NuvoAir & FDA 510 (K) & Manage \\ \cline{2-6} 
 & Cancer & Oleena~\cite{Oleena:2020} & Voluntis & FDA 510 (K) & Manage \\ \cline{2-6} 
 & \multirow{4}{*}{Diabetes} & BlueStar~\cite{Quinn:2011} & WellDoc & FDA 510 (K) & Manage \\ \cline{3-6} 
 &  & Insulia~\cite{Insulia:2020} & Voluntis & FDA 510 (K) & Manage \\ \cline{3-6} 
 &  & Transform~\cite{Alwashmi:2019} & Virgin Pulse & CDC DPP & Manage \\ \cline{3-6}
 &  & Virta~\cite{Athinarayanan:2019} & Virta Health & \begin{tabular}[c]{@{}c@{}}-\end{tabular} & Manage \\ \cline{2-6} 
 & \multirow{3}{*}{\begin{tabular}[l]{@{}l@{}}Diabetes,\\Hypertension\end{tabular}} & \begin{tabular}[l]{@{}l@{}}Dario~\cite{dariohealth_2020}\end{tabular} & DarioHealth & FDA 510 (K) & Manage \\ \cline{3-6} 
 &  & Omada~\cite{Sepah:2017} & Omada Health & CDC DPP & Manage \\ \cline{3-6} 
 &  & Livongo~\cite{Downing:2016} & Teladoc Health & FDA 510 (K) & Manage \\ \cline{2-6}
 & HIV & PositiveLinks~\cite{PositiveLinks} & PositiveLinks Platform & - & Manage \\ \cline{2-6}
 & \multirow{3}{*}{\begin{tabular}[l]{@{}l@{}}Stroke, \\Neurological \\disorders \end{tabular}} & \begin{tabular}[l]{@{}l@{}}Constant Therapy~\cite{constanttherapy}\end{tabular} & Constant Therapy Health & FDA Breakthrough & Manage \\ \cline{3-6} 
 &  & NeuroEyeCoach~\cite{neuroeyecoach} & NovaVision & FDA 510 (K) & Manage \\ \cline{3-6} 
 &  & Nirvana~\cite{nirvanavr} & BTS Bioengineering & - & Manage \\
 \hline 
\end{tabular}}
\end{table*}

\subsection{DTx Regulations} 

The development of the digital healthcare sector has led to the emergence of new software concepts such as software as a medical device (SaMD). The International Medical Device Regulator Forum, which is a coalition of regulators from various countries, defines SaMD as ``software intended to be used for one or more medical purposes that perform these purposes without being part of a hardware medical device''~\cite{IMDRFSaMD}. In general, SaMD includes DTx as well as other clinical decision-making software such as computer-aided diagnosis for radiologists. Similarly to hardware-based medical devices, SaMDs have highlighted the need for regulatory oversight and approvals. the US FDA requires SaMD to receive approval by demonstrating safety and effectiveness either through the Premarket Notification 510 (K) process of comparison with \emph{similar} legally marketed devices (known as substantial equivalence proof)~\cite{FDA-501k}, or the De Novo classification process if a comparison is not feasible (new technology)~\cite{FDA-DeNovo}. In 2018, US FDA further introduced breakthrough designations for SaMDs with more effective treatment or diagnosis of life-threatening human diseases such that manufacturers can interact with the FDA's experts to expedite such review processes~\cite{Johnston2020BDP}. 

There is also a newly created pre-certification program that certifies SaMD ``developers'' instead of SaMD ``products'' so that the total product lifecycle can be managed (e.g., product quality, patient safety, and cybersecurity)~\cite{FDA-PreCert}. 

To date, most applications and services, classified as DTx, have received 510 (K) clearance by submitting RCT results from users with new technology and those with make-believe technology as a placebo to demonstrate their effectiveness. WellDoc's BlueStar diabetes management system was one of the first FDA 510 (K) submissions approved in January 2017. Pear Therapeutics reSET (drug addiction) and Akili's EndeavorRx (attention deficit) were approved via the De Novo process and validated by comparing existing treatment methods with and without the newly developed techniques~\cite{kollins2020novel, FDA2016}. 

Besides the US FDA approvals, for diabetes management, there is another certification called the full Centers for Disease Control and Prevention (CDC) recognition for Diabetes Prevention Program (DPP). Several commercial DTx products have received full CDC recognition for DPP, such as Noom~\cite{noom2017} and Omada~\cite{omada2018}, which provide users with diabetes programs based on computerized cognitive behavioral therapy. They were recognized for the effectiveness of diabetes management through clinical trials for participants in the National DPP program, which lasted more than 16 weeks~\cite{Sepah:2017,michaelides2016weight}.

\begin{table*}[]
\centering
\caption{Digital Therapeutics with JITAI}
\label{jitai}
\begin{tabular}{clll}
\hline
\textbf{\begin{tabular}[c]{@{}c@{}}App name\\ (target disease)\end{tabular}}    & \multicolumn{1}{c}{\textbf{JITAI components}} & \multicolumn{1}{c}{\textbf{Contents}}   & \multicolumn{1}{c}{\textbf{Data types}} \\ \hline
\multirow{8}{*}{\begin{tabular}[c]{@{}c@{}}A-CHESS\\ (alcoholism)\\~\cite{Gustafson2014ACHESS}\end{tabular}} & Distal outcome  & \begin{tabular}[c]{@{}l@{}}Regular reduction on the number of risky drinking days (4, 8, and 12 months after discharge)\end{tabular}& numeric   \\ \cline{2-4} 
& Proximal outcome& \begin{tabular}[c]{@{}l@{}}Weekly Brief Alcohol Monitoring (BAM) index result\end{tabular}& numeric   \\ \cline{2-4} 
& Tailoring variables  & Location of the individual    & object    \\ \cline{2-4} 
& Intervention options & \begin{tabular}[c]{@{}l@{}}Helping patients stay sober by prompting them if they go near the high risk places\end{tabular}    & string    \\ \cline{2-4} 
& \multirow{2}{*}{Decision points}    & \begin{tabular}[c]{@{}l@{}}The moment when the user gets closer to high-risk location\end{tabular}& timestamp \\ \cline{3-4} 
&  & When the user presses the panic button  & timestamp \\ \cline{2-4} 
& \multirow{2}{*}{Decision rules}& \begin{tabular}[c]{@{}l@{}}Whether the user gets closer to the high-risk location or not\end{tabular}& boolean   \\ \cline{3-4} 
&  & Whether the panic button is pressed or not   & boolean   \\ \hline
\multirow{8}{*}{\begin{tabular}[c]{@{}c@{}}Q-Sense\\ (smoking)\\~\cite{naughton2016context}\end{tabular}}    & Distal outcome  & Smoking abstinence  & boolean   \\ \cline{2-4} 
& Proximal outcome& Reduced stays within high risk locations   & numeric   \\ \cline{2-4} 
& \multirow{2}{*}{Tailoring variables}& Location of the individual
  & object    \\ \cline{3-4} 
&  & \begin{tabular}[c]{@{}l@{}}User self-reports (i.e., mood, stress, urge, current context, and whether other smokers \\are present) collected when the user starts smoking during the pre-quit phase\end{tabular} & mixed\\ \cline{2-4} 
& \multirow{2}{*}{Intervention options}    & \begin{tabular}[c]{@{}l@{}}Tailored support message (prefilled based on the user's demographics and smoking survey)\end{tabular}& string    \\ \cline{3-4} 
&  & Feedback messages based on the user's smoking reports & string    \\ \cline{2-4} 
& Decision points & \begin{tabular}[c]{@{}l@{}}When the user enters and stays in the high risk location for more than 5 minutes \end{tabular} & timestamp \\ \cline{2-4} 
& Decision rules  & \begin{tabular}[c]{@{}l@{}}Whether the user enters and stays in the high risk location for more than 5 minutes \end{tabular}   & boolean   \\ \hline
\multirow{6}{*}{\begin{tabular}[c]{@{}c@{}}DietAlert\\ (obesity)\\\cite{Goldstein17Return}\end{tabular}}  & Distal outcome  & Weight loss    & numeric   \\ \cline{2-4} 
& Proximal outcome& If exceeding calorie limit or not  & boolean   \\ \cline{2-4} 
& Tailoring variables  & \begin{tabular}[c]{@{}l@{}}21 tailoring variables (temptation, boredom, hunger, and planning food intake, exercise, etc.)\end{tabular}  & numeric   \\ \cline{2-4} 
& Intervention options & \begin{tabular}[c]{@{}l@{}}157 subdivided elements provided (e.g., think of something fun to try right now)\end{tabular}   & string    \\ \cline{2-4} 
& Decision points & When the user replies to an EMA prompt  & timestamp \\ \cline{2-4} 
& Decision rules  & \begin{tabular}[c]{@{}l@{}}Use supervised machine learning to identify risk for lapse occurrence\end{tabular}& numeric   \\ \hline
\end{tabular}
\end{table*}

\section{Review of Recent DTx Therapies}

Digital therapeutics apps use a variety of behavior change techniques such as feedback on behavior and goal setting, social support, instructional guidelines, and self-monitoring of behavior, and outcome(s) of the behavior. As DTx apps are not yet well established, to grasp the trends, a non-probability sampling method---snowball sampling---was performed. Potential DTx apps were identified online based on our judgment of mental and physical disorder categories until the required sample size was reached. A solid effort was made to embrace most DTx apps introduced/released in the last three years, as well as those with FDA certificates (i.e., De Novo and 510 (K)) and recognition by the CDC.
This paper may not exhaustively review all relevant literature, but it clearly reflects the state-of-the-art technology of DTx apps. The insights obtained from the process are summarized according to the therapeutic areas and diseases they target. For each DTx, regulatory endorsement is checked (e.g., 510 (K), De Novo, and CDC full recognition), and intended use is categorized based on Digital Therapeutics Alliance~\cite{dta2018dtx} categories: (1) management or prevention of a medical disorder/disease, (2) medication optimization, and (3) medical disease or disorder treatment.
 
As shown in Table.~\ref{dtx}, the established DTx apps span two therapeutic areas, namely, mental and physical disorders. The mental disorders include ADHD~\cite{kollins2020novel}, depression~\cite{Meyer:2009,Fitzpatrick:2017}, insomnia~\cite{Ritterband:2009}, panic disorder~\cite{Tolin:2017}, and substance abuse~\cite{Campbell:2014,Christensen:2014}. Among DTx apps in this area, Pear Therapeutics's reSET~\cite{Campbell:2014} is the first prescription DTx product (i.e., De Novo) designed to deliver behavioral therapy to treat substance use disorder; reSET-O~\cite{Christensen:2014} is another prescription DTx product designed to deliver cognitive behavioral therapy for opioid use disorder. As a research prototype, EndeavorRx~\cite{kollins2020novel} is a home-based, video game-like digital app for the treatment of inattention and cognitive dysfunction in pediatric patients with ADHD. 

Another major branch of DTx apps is physical disorders, such as asthma~\cite{Merchant:2016}, cancer~\cite{Oleena:2020}, diabetes~\cite{Quinn:2011,Insulia:2020,Athinarayanan:2019}, diabetes/hypertensions~\cite{dariohealth_2020,Sepah:2017,Downing:2016}, and diabetes/obesity~\cite{Alwashmi:2019}. Among the DTx apps in this area, BlueStar~\cite{Quinn:2011}, a digital therapeutic app developed by WellDoc, has assisted patients and providers in improving glucose control by using real-time data and feedback to support healthy behaviors, such as medication adherence, diet and exercise control, and psychosocial wellness. BlueStar has demonstrated the capacity to shift HbA1c levels in populations with diabetes. Interestingly, the Dario Blood Glucose Monitoring System~\cite{dario_2020} by DarioHealth, Omada~\cite{omada_hyper} by Omada Health and Livongo~\cite{livongo:2018} by Teladoc Health combine the DPP with hypertension, as 80\% of people with type 2 diabetes also have high blood pressure. Both hypertension and diabetes result from metabolic syndromes. Thus, they develop sequentially in the same individual~\cite{hypertension,cheung2012diabetes}.

DTx apps can incorporate \emph{just-in-time support} which is an attempt to provide the right type (or amount) of support at the right time (i.e., neither too early nor too late). For example, the ongoing monitoring of an individual with mobile sensing can identify when these events/conditions occur (i.e., when support is needed). As an emerging technology-driven, behavior change-oriented intervention type, just-in-time adaptive intervention (JITAI) capitalizes on real-time sensor data collected via mobile sensing technology (e.g., smartphones and wearables) to adaptively trigger appropriate support in situ~\cite{Nahum-Shani}. The major components of DTx with JITAI can be summarized as follows:

\begin{itemize}
\item \textit{Distal outcome:} the long-term goal (\ie behavior change) of a DTx app (e.g., reduction of sedentary behavior in older adults~\cite{Thomas20015JITAI}). 
\item \textit{Proximal outcome:} the short-term goal of a DTx product, which mediates the effect of the intervention on the distal outcome. In other words, the proximal outcome ensures progress toward the distal outcome (e.g., increased number of active breaks from prolonged sitting over a day).
\item \textit{Tailoring variables:} the collection of behavioral and contextual data that can identify when behavioral support might be most effective (e.g., accumulated sitting time, location of individual, time of day, response to support prompts sent earlier, and frequency of support prompts). 		 
\item \textit{Intervention options:} adequate intervention
options (e.g, suggestions of light movements, positive feedback, and encouragements to repeat the light movements) that are delivered to the user when an opportune moment for behavior change support is detected.
\item \textit{Decision points:} a marked moment when a decision to send an intervention option is made considering tailoring variables (e.g., the time between 5 pm and 9 pm).
\item \textit{Decision rules:} systematic rules of whether to provide certain intervention options based on past intervention options and tailoring variables (e.g., location-based feedback is provided if a user becomes sedentary for more than 30 minutes). 
\end{itemize}

\begin{figure*}[!t]
\centering
\includegraphics[width=\textwidth]{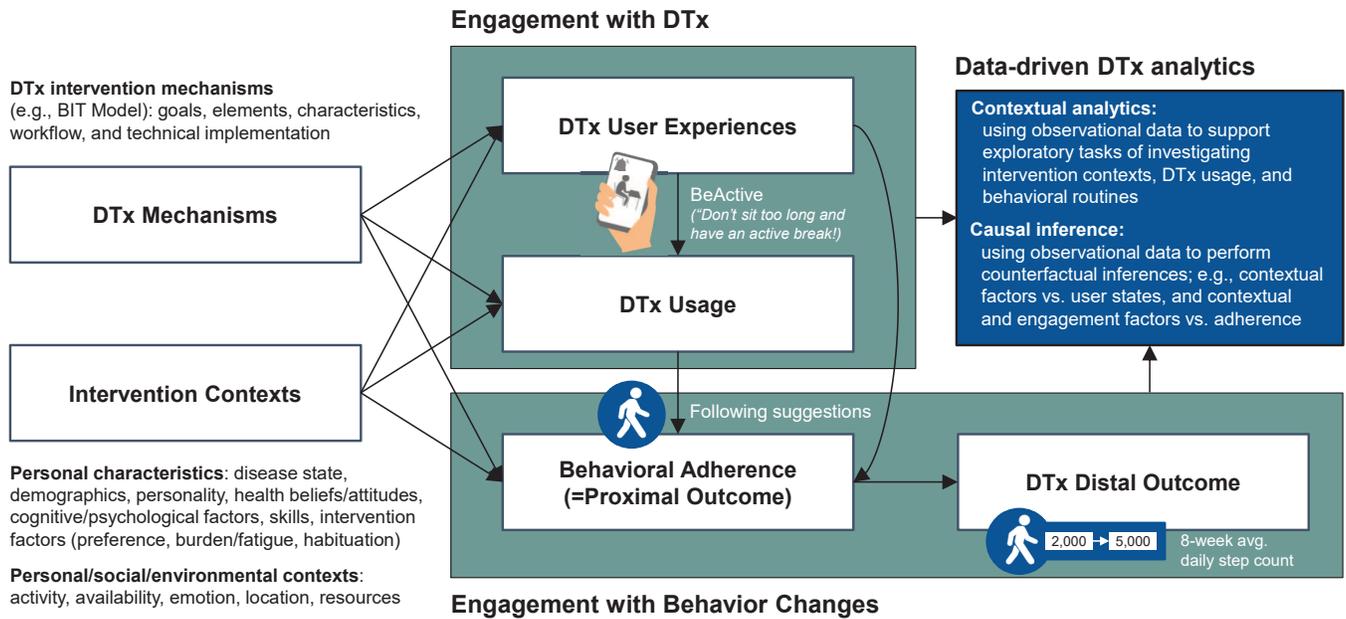}
\caption{Data-driven DTx analytics framework }
\label{fig:dtxframework}
\end{figure*}

JITAI-based DTx apps were further reviewed in the context of JITAI with relevant data types in Table.~\ref{jitai}, which describes DTx apps that target alcohol addiction~\cite{Gustafson2014ACHESS}, tobacco addiction~\cite{naughton2016context}, and obesity~\cite{Goldstein17Return}.

\emph{Alcohol addiction}: A-CHESS~\cite{Gustafson2014ACHESS} offers JITAIs to reduce risky drinking days (i.e., distal outcome) for patients with alcohol use disorder after leaving a treatment facility. A-CHESS monitors data collected via mobile sensing to trigger just-in-time interventions in the form of messages and tracks the reduced brief alcohol monitoring index (i.e., proximal outcomes) weekly to make progress toward the distal outcome. The app uses tailoring variables (e.g., location of an individual, user data submissions, messages sent/received via the app) and intervention options such as sending a prompt notice if the user comes near places associated with a high risk for drinking. A-CHESS defines the moment when the user's location is close to the high-drinking-risk place and the moment when the panic button is pressed as decision rules and decision points.

\emph{Cigarette smoking}: JITAI for smoking cessation can leverage self-reports and automatic sensing for risk assessment. Smart-T~\cite{businelle2016ecological, hebert2020mobile} estimated the risk of smoking using the user's self-report data via ecological momentary assessment (EMA) (e.g., urge to smoke, stress, alcohol consumption, and availability of cigarettes). Sense2Stop~\cite{spring2000sense2stop} leverages automatic stress sensing mechanisms to deliver support messages when high-stress moments are detected. Q-Sense~\cite{naughton2016context} learns a user's smoking behavior, emotional conditions, and places via EMA, and the user's high-risk locations are tracked to deliver just-in-time support messages. 
Q-Sense leverages user self-assessment and mobile sensing to trigger just-in-time support messages for smoking abstinence (i.e., distal outcome). It also tracks proximal outcomes (i.e., not staying in a high-risk location) to make progress toward the distal outcome. Self-assessment and location are used as tailoring variables to trigger the intervention. Q-Sense checks whether a user stays in high-risk locations for longer than 5 min (known as a geofensing technique), which is considered a decision rule, and the current timestamp becomes the decision point. Intervention options include tailored support and feedback messages. Tailored support messages are pre-populated based on the user's demographics and smoking survey. Feedback messages are sent based on the smoking behavior pattern of the user and on user self-reports (e.g.,``Based on 12 reports, did you know you smoke 25\% of the time when working?''). 

\emph{Obesity}: DietAlert~\cite{Goldstein17Return} focuses on lapses in a weight-control diet among overweight and obese individuals. For just-in-time intervention, the app tracks whether the user exceeds the calorie limit as the proximal outcome. DietAlert tracks 21 tailoring variables (e.g., temptation, boredom, hunger, and exercise) and 157 subdivided elements (e.g., thinking of something fun for boredom). DietAlert uses supervised machine learning to identify lapse occurrence risk as a decision rule and keeps track of the user responses for decision points (six semi-random time intervals throughout the day, spaced approximately 2--3 h apart).

\section{Overview of Data-Driven DTx Analytics}

This section provides an overview of the core elements of data-driven DTx analytics (Fig. \ref{fig:dtxframework}). The major flow of data-driven DTx analytics involves DTx user experience, DTx usage, and behavior adherence (proximal and distal outcomes). These elements are closely related to the offered DTx mechanisms (e.g., goals and strategies) and users' intervention contexts (e.g., user traits and environments). In addition to traditional clinical trials with RCTs, we discuss how sensor and interaction data can be used to enable contextual and causal analytics.

\emph{DTx engagement}: We used the existing conceptual model \emph{engagement} by Yardley et al.~\cite{Yardley2016Understanding}, which made distinctions between engagement with technological and behavioral aspects. Engagement with technological aspects (i.e., DTx user experience and DTx usage) refers to how a user makes use of software for behavior changes. In contrast, engagement with behavioral aspects refers to how a user initiates and sustains behavior changes, which can possibly lead to ``positive outcomes'' (e.g., reaching weight loss goals). Intervention software plays an important role in scaffolding behavior changes, so that sustained usage of intervention software is no longer necessary for achieving ``positive outcomes.'' In other words, if a user has finished mastering a skill offered by the intervention software, behavioral changes can be successfully maintained even without software usage. This distinction was also used in Alshurafa et al.'s work~\cite{Alshurafa18IsMore}, where engagement with intervention software referred to DTx usage (e.g., duration and frequency of app usage) and engagement with behavior changes referred to \emph{adherence to the intervention} (e.g., proximal outcomes). In this work, ``engagement'' refers to DTx software usage and user experience, and ``adherence'' refers to behavior changes prompted by intervention. In Section \ref{subsec:datadrivenmodels}, the modeling of engagement and adherence using datasets is further discussed.

\emph{DTx mechanisms}: Existing intervention models include the behavior intervention technology (BIT) model~\cite{mohr2014behavioral} and Fogg's model~\cite{fogg2009creating}. The BIT model is a software-based intervention at both theoretical and technical levels: (1) theoretical aspects include intervention goals (e.g., weight loss and healthy eating) and behavioral intervention strategies (e.g., goal setting, education, monitoring, and feedback), and (2) technical aspects include behavioral intervention technology elements (e.g., information delivery, data collection, and reports), technical characteristics (e.g., medium, complexity, aesthetics, and personalization), and workflow of the intervention to determine when to deliver the intervention to the user (e.g., context-based scheduling). 

\emph{Intervention contexts}: These include both personal characteristics and contextual factors. An existing behavior change model for Internet interventions~\cite{Ritterband:2009BCM} considers (1) user characteristics (e.g., disease states, demographics, traits, beliefs and attitudes, physiological factors, and skills), (2) intervention experiences (e.g., user preferences, perceived burdens/fatigue, and habituation), and (3) intervention environments (e.g., personal, professional, and community aspects). This ``contextual'' model can be further extended by using existing context representation models for ubiquitous computing where mobile, wearable, and Internet-of-Things (IoT) devices are used to enable context-aware services~\cite{Schmidt1999Context, Schilit1994Context}. Additional contextual information includes both human factors (e.g., users' cognitive and affective states, social environment, and tasks) and physical environments (e.g., location/place, lighting condition, and temperature). Intervention context data can be collected via user self-reports and automatic inferences from sensor data collected passively via machine learning.
 
\emph{Data-driven DTx analytics}: An RCT, as the gold standard for clinical evaluation, helps to evaluate the efficacy of a newly developed digital intervention, as opposed to the traditional approach. However, it is difficult for RCTs to identify the key contextual patterns and intervention factors that influence efficacy. Sensor and interaction data passively collected during RCTs can be used in contextual and causal analytics. Contextual analytics supports exploratory tasks of understanding contextual factors related to DTx usage such as lifestyle (behavioral routines), intervention contexts, and psychological states of the user (e.g., emotion and stress) over the intervention period. Causal analytics (counterfactual inference) helps to investigate how individual intervention elements and user engagement influence behavioral adherence (i.e., proximal and distal outcomes).

\section{Key Components of Data-Driven DTx Analytics Flow}

The overall flow of data-driven DTx analytics is shown in Fig. \ref{fig:dtxprocess}. Conducting large-scale RCTs is very expensive and time-consuming and constitutes a major bottleneck in rapid software iteration. The goal of data-driven DTx analytics is not only to evaluate the efficacy of the intervention, but also to acquire design and clinical insights through a comprehensive analysis of the passively collected sensor and interaction data from DTx field trials. The major components of a DTx analysis flow include (1) data collection and user management in field trials, (2) data-driven models for engagement and adherence, (3) experimental design and contextual/causal analytics, (4) data visualization, and (5) DTx design and development processes, which are reviewed in detail as follows. 

\begin{table*}[!t]
\centering
\caption{Comparing major features of existing CTMS systems}
\label{CTMS_table}
\scalebox{0.8}{
\begin{tabular}{lcccccccc}
\hline
\multirow{2}{*}{\textbf{\begin{tabular}[c]{@{}c@{}}CTMS \end{tabular}}} & \multicolumn{8}{c}{\textbf{Major features supported}} \\ \cline{2-9} 
 & \textbf{Milestone mgmt} & \textbf{Education mgmt} & \textbf{Resource mgmt} & \textbf{Subject mgmt} & \textbf{Data collection} & \textbf{Data analytics} & \textbf{Legal service} & \textbf{Financial service} \\ \hline \hline
ArisGlobal~\cite{arisglobal_2020} & O & O & O & O & O & O & O & O \\
BioClinica~\cite{bioclinica_2020} & O & O & O & O & O & O & O & O \\
Bio-Optronics~\cite{bio-optronics_2020} & O & O & O & O & O &  & O & O \\
DSG~\cite{dsg_2020} & O & O & O & O & O &  & O & O \\
eResearch Technology~\cite{bioclinica_2020} & O & O & O & O & O & O & O &  \\
MedNet Solutions~\cite{mednet_2020} & O & O & O & O & O & O & O & O \\
Medidata Solutions~\cite{medidata_2020} & O & O & O & O & O & O & O & O \\
PAREXEL International~\cite{parexel_2020} & O & O & O & O & O & O & O & O \\
Trial By Fire Solutions~\cite{simpletrials_2020} & O &  &  & O & O &  & O & O \\
Veeva Systems~\cite{veevasystems_2020} & O & O &  & O & O & O & O & O \\ \hline
\multicolumn{9}{l}{O: Supported}
\end{tabular}}
\end{table*}

\subsection{Data Collection and User Management}

We review technologies enabling data collection and user management, such as clinical trial management systems and general-purpose platforms. Data sources are largely based on sensor and interaction data from smart devices (e.g., phones, wearables, and IoT devices); thus, smart devices and their interactions are discussed.

\subsubsection{Clinical Trial Management System}

As stakeholders perform various clinical trials with a certain level of quality, it is necessary to utilize an information technology system to support data collection and user management in standardized clinical trial processes, to comply with relevant regulations. Based on a commonly agreed upon standardized process, a clinical trial management system (CTMS) was designed and developed as a comprehensive system to comply with privacy and security regulations. Table.~\ref{CTMS_table} summarizes the representative CTMSs ~\cite{CTMS2019} and states the eight key supporting features required to conduct standardized clinical trials, as discussed in \cite{Park2018CTMS}. These features include (1) milestone management (e.g., managing timeline of clinical trial such as regulatory completion), (2) education management (e.g., training of researchers of the study by CTMS expertise team or providing educational services to ensure that subject remains in the study), (3) resource management (e.g., managing medical devices and clinical drug import and export), (4) subject management (e.g., managing the health information of the subject and subject recruitment and scheduling), (5) data collection (e.g., using electronic data capture (EDC) system to collect clinical data in electronic format for human clinical trials instead of the traditional paper-based data collection methodology, to streamline data collection, and to expedite the time for marketing the drugs and medical devices), (6) data analytics (e.g., providing visualization of clinical research data using dashboard or analyzing clinical data using other analytic modules), (7) regulatory services (e.g., adhering to guidance and directives from the regulatory agencies such as FDA and European Medicines Agency), and (8) financial services (e.g., financial management of clinical trials).

Ten well-known CTMSs reported in a recent study~\cite{CTMS2019} were selected for this review. Product brochures, publicly available videos, and white papers were scrutinized to check whether the eight essential features were supported. Most systems fully support the key features; however, several systems do not provide some of the features, possibly because their focus is on specific customer segments. It is interesting to note that Medidata Solutions~\cite{medidata_2020} and Parexel International~\cite{parexel_2020} provide additional EDC support to collect sensor data (e.g., pulse oximeters, blood pressure meters, activity trackers, glucometers, spirometers, body weight scales, and ECG) and wearables. The other CTMSs do not closely match these standards. Major players include IBM Watson Health~\cite{ibm_watson_health}, Oracle~\cite{oracle_2020}, Nextrials~\cite{nextrials_2020}, and Winchester Business Systems~\cite{winchester_2020}.

\begin{figure*}[!h]
\centering
\includegraphics[width=0.9\textwidth]{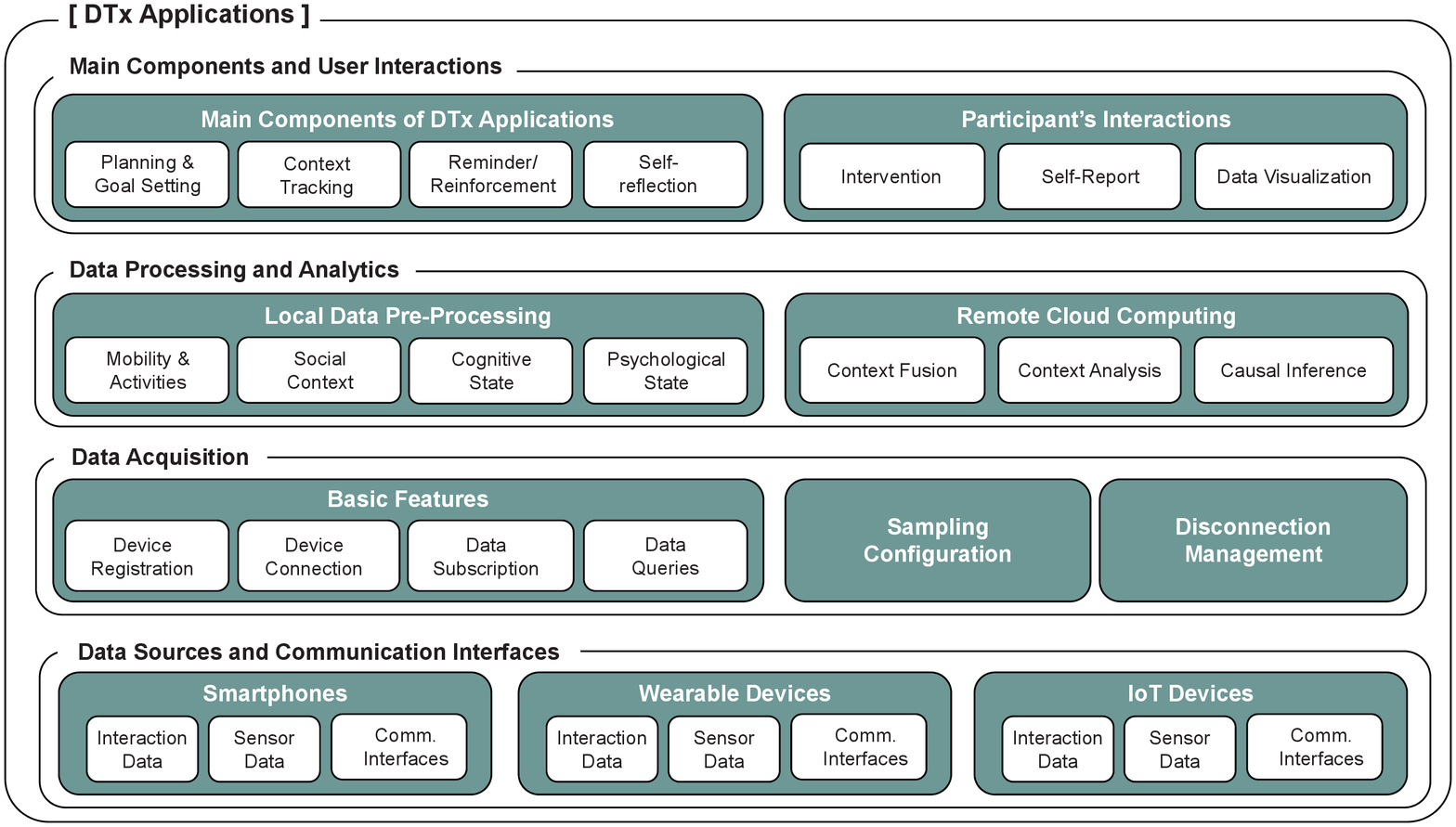}
\caption{The architecture of data-driven DTx analytics platform}
\label{fig:platform}
\end{figure*}

\subsubsection{General-Purpose Mobile Sensing Platforms}

Compared with CTMS, research communities in mobile and ubiquitous computing have made intensive efforts to build general-purpose platforms to collect sensor data from mobile and wearable devices (e.g., smartwatches and smartbands). As shown in Fig.~\ref{fig:platform}, here, the platform components are conceptualized with a layered architecture. (1) \emph{Data source and Communication interfaces} layer includes the collection of sensor data (e.g., accelerometers, gyroscopes, magnetometers, and GPS) from smartphones and other wirelessly connected devices (e.g., smartwatches, smartbands, and IoT devices). Through a communication interface (i.e., Bluetooth, WiFi), data from the devices are transferred to smartphones and stored locally. (2) \emph{Data acquisition} layer includes the implementation of an agent that performs basic functions (e.g., device registration, connection, subscription, and data queries), sampling rate adjustment for sensing, and connection failure management (e.g., wireless connectivity failures between a wearable and smartphone). The obtained data are then archived into a local database. (3) \emph{Data processing and analytics} layer implements local data preprocessing and remote cloud computing for context analytics and causal inference. Various machine learning algorithms can be executed on either local devices (e.g., smartphones or local servers) or remote servers. Lightweight machine learning algorithms on edge devices process raw sensor data to extract contextual features and behavioral markers (e.g., mobility features, physical activities, and users' cognitive and psychological states). Computationally intensive analytics can be performed via remote cloud computing infrastructure. For example, causal inference helps to establish causality of the effect of the provided DTx treatments, and contextual analytics helps to understand users' intervention contexts, such as problematic situations and the influence of user contexts on adherence. (4) \emph{DTx} layer includes the main components of DTx applications and user interactions. DTx applications (e.g., smoking cessation and depression management) usually contain pipe-lined procedures (i.e., planning and goal setting, context tracking, reminding/reinforcement, and self-reflection). The participant's interactions involve behavioral interventions, user self-reporting to the platform, and visualization of the user's daily (or longer-term) summary toward the goal. A more detailed review on DTx mechanisms can be found in Section V-E.

General-purpose platforms mainly focus on providing reliable and scalable sensor data collection, and therefore, compared to CTMS, offer only a limited number of functionalities, such as research resource/progress management and logistics support. Recently, several well-known data-collection platforms have been introduced, including OpenDataKit~\cite{opendatakit}, AWARE~\cite{ferreira2015aware}, and mCerebrum~\cite{hossain2017mcerebrum}, which enable researchers to collect user self-reports and stream high-frequency sensor data to the remote cloud for storage and data processing. OpenDataKit adopts a middleware design approach that allows third-party app developers to minimize their efforts in sensor-specific codes via reusable sensor drivers. For example, it is possible to download new sensor capabilities from the application market and use them without any need for modifications. OpenDataKit further manages the discovery, communication channels, and data buffers for extensibility. AWARE provides mobile data-logging tools and supports external sensor plugins for the collection and abstraction of sensor data for context-aware service delivery. System scalability is an important issue for these platforms. mCerebrum significantly improves the scalability of storage for high-rate sensor data and provides several fine-tuned features, such as sensor duty cycling, energy-optimized context inference computation as a shared service, and sensor data quality assessment. AWARE and mCerebrum can deliver local interventions, but the cloud components of these frameworks are limited to automatically triggered interventions based on past user data. Data collection requires sustained user engagement. SARA~\cite{rabbi2017sara} integrates gamified engagement strategies, including contingent rewards, badges for completing active health tasks, funny memes/gifs and life insights, and health-related reminders or notifications to incentivize data collection.

There are other data collection platforms such as the Personal Health Intervention Toolkit (PHIT)~\cite{eckhoff2015platform} that exclusively target health interventions. PHIT is a framework that facilitates mobile data collection and supports health interventions for research purposes. PHIT allows researchers to explore and perform health status analysis, intervention recommendations, and self-help activities, and to explore uploaded data through a Web-based portal. The major departure of PHIT from CTMS and general-purpose data collection platforms is that it provides an intelligent virtual advisor that analyzes real-time data from devices and facilitates tailored interventions. A recent study proposed Duty~\cite{kizakevich2012phit,kizakevich2018phit}, which integrates mindfulness-based relaxation, behavioral education in sleep quality and alcohol use, and psychometric and psychophysiological data capture. Duty used PHIT as a personalized health intervention framework for acquiring data, including self-report instruments, EMA diaries, cognitive tests, and game-like activities.

\subsubsection{Smart Devices and User Interactions}

Current commercial DTx apps (e.g., Woebot and reSET) are mostly in software format using off-the-shelf smartphones, but they also leverage wearable trackers (e.g., Fitbit and Apple Watch) and IoT devices (e.g., Propeller Health's Bluetooth inhalers and Insulia's Blood Glucose Meters). Mobile, wearable, and IoT technologies support fine-grained sensing and tracking of users' states ranging from physiological signals, such as heart rates and skin temperature, to physical activities, social interactions, and user's interactions with digital devices (also known as human-computer or human-machine interactions). These smart devices provide \emph{physical actuation} (e.g., controlling light bulbs or thermometers) or \emph{virtual actuation} (e.g., launching apps and sending emails). Beyond local physical sensing, it is possible to use Web-based sensor data such as weather and air quality data via open application programming interfaces (APIs) as virtual sensors~\cite{Nakazawa15Sensorizer}.

Existing DTx apps that consider just-in-time support often use diverse real-time sensing features such as activity tracking. For example, the BeActive system continuously monitors a user's physical activity and provides just-in-time feedback when the user becomes sedentary for more than 50 min~\cite{sun2020beactive}. This type of real-time monitoring can be supported by the use of diverse sensors. Current off-the-shelf smartphones include various sensors, such as microphones, GPS, motion sensors, a compass, a light sensor, and cameras. Sensor data can be passively collected in the background (known as opportunistic sensing; e.g., tracking a user's location traces), or users are asked to perform specific tasks for data collection (known as participatory sensing; e.g., taking a photo for food journaling)~\cite{Lane10}. In particular, sensor and interaction data from smartphones or smartwatches facilitate a fine-grained understanding of user contexts and the detection of various events of interest, such as activity tracking with motion sensors and social interaction tracking with audio sensing or call/SMS log tracking. Wearable devices such as smartwatches and activity trackers offer a similar level of sensing, but the key departure is their support for sensing physiological signals, such as heart rate and skin temperature, which are useful for detecting stress and emotions (see Cowley et al.'s review~\cite{Cowley16}). Consumer electronics offer only limited capabilities for data collection, whereas wearable sensors for research purposes (e.g., Empatica E4 and Shimmer3) provide APIs for accessing raw data; however, their cost is an order of magnitude greater.

IoT devices can be classified based on their functionality and embeddedness. The major functionalities of IoT devices in domestic and office environments include IoT control (e.g., hubs or voice assistants), actuation (e.g., lighting and switches), and sensors (e.g., motion, temperature, and air quality). Most IoT devices are standalone products so that users can wirelessly control the connected door locks and thermostats through the Internet. Recent smart appliances, such as air conditioners, refrigerators, and air purifiers, are equipped with smart features such as state monitoring (e.g., operating condition logging and environment states) and remote control (e.g., temperature control). Unlike human activity monitoring with standalone IoT devices, smart appliances naturally come with sensors. As human appliance interactions are naturally part of activities of daily living, they provide useful information for health behavior monitoring (e.g., cognitive decline tracking~\cite{Lee2011PillsPhones}). Furthermore, smart appliances can help closely monitor home environments such as room temperature and air quality states (e.g., \ch{CO2}, PM2.5, and VOC). IoT devices or IoT-enabled appliances are typically connected to central hubs for integrated control over the Internet. Voice assistants include Amazon Echo and Google Home, which provide natural language support for information activities (e.g., Q\&A) and device control (e.g., turning off bulbs). The type of IoT device control offered by IoT hubs, such as SmartThings Hub, provides a novel means of enabling DTx; for example, IoT devices can offer context-aware intervention for personalized sleep education (e.g., automatically turning off lights)~\cite{Lee2017SelfExperimentation}.

\subsection{Data-Driven Models for Engagement and Adherence} 
\label{subsec:datadrivenmodels}

The digital intervention and passive sensor data are used to measure \emph{user engagement in digital intervention} (or software usage) and \emph{user adherence to behavioral changes}. Compared to ``software engagement,'' adherence to behavioral changes is relatively easy to measure by tracking the user behavior based on user interactions (e.g., whether the content is consumed?) and passive sensing (e.g., whether physical activity has happened?) or by asking users to self-report. Diverse data-driven models for user engagement are discussed next; reviewing such models provides insight into the types of data that can be collected for data-driven DTx analytics. 

A simple approach to measuring user engagement (and behavioral adherence) involves collecting self-reported feedback (e.g., level of satisfaction)~\cite{Taki2017Assessing}. Self-reports can be submitted at the end of experiments using well-known user engagement and usability scales, such as the system usability scale~\cite{SUS1986}, usefulness, satisfaction, and ease of use questionnaire (USE)~\cite{USE2001}, and the user engagement scale (UES)~\cite{OBrien2018UES}. These metrics consider diverse dimensions; for example, UES considers attention, usability, aesthetics, endurability, novelty, and involvement, and USE considers usefulness, ease of use, ease of learning, and satisfaction. In the case of digital health apps, information quality, trust, and security are also important~\cite{Stoyanov2015MARS}. While users typically answer these kinds of survey questionnaires at the end of an intervention period, it is also possible for participants to report their experiences on a shorter time scale (e.g., whenever some events happen or once a day) via EMA. An in-depth understanding of user engagement is difficult to acquire from surveys; in this case, follow-up interviews and thematic analysis of the interview data are required~\cite{Brhel2015Exploring}. 

Prior studies examined various usage metrics of digital interventions~\cite{Alshurafa18IsMore, Taki2017Assessing, BenZeev2016mHealth}. Basic usage metrics include the frequency of app use, user responses to proactive prompts, and on-demand usage that can be aggregated over a specific time window (e.g., per day or per week)~\cite{BenZeev2016mHealth}. In addition, interaction-level metrics can be considered. For a given use instance, the frequency of fine-grained interaction types, such as glancing, checking, and brief/detailed reviewing and their durations~\cite{Alshurafa18IsMore} can be summarized. In-depth interaction patterns can also be analyzed, such as the inter-checking intervals and receptivity to push notifications~\cite{Taki2017Assessing}. It is also possible to build a composite engagement index, a simple approach for a linear weighted sum of usage metrics~\cite{Taki2017Assessing} such as the number of pages viewed per session, frequency of app access, receptivity to push notifications, inter-checking intervals, and self-reported feedback (satisfaction).

In addition to such usage metrics, prior studies have examined the use of passive sensor data. Traditional engagement models based on flow theory mainly consider user allocation of attentional resources and their affective states. It is possible to measure a user's attention via behavioral and physiological sensor data, such as eye-tracking, facial expression, mouse movements, electrodermal activity (EDA), ECG, and electroencephalogram (EEG) data. Martey et al.~\cite{Martey2014Measuring} analyzed how passive sensor data (e.g., EDA and mouse movements/clicks) can be used to estimate self-reported engagement. Lascio et al.~\cite{Lascio2018EDA} explored how EDA data can be used to model a student's emotional engagement (i.e., enthusiasm and enjoyment) with the support vector machine (SVM). Mathur et al.~\cite{Mathur2016Engagement} used EEG data to model user engagement from mobile context data and identified a strong correlation between automatically detected engagement scores and users' subjective perception of engagement. Pielot et al.~\cite{Pielot2017Beyond} considered consumption of recommended content as ``engagement'' and used machine learning (i.e., XGBoost) to identify latent engagement factors, which are based on diverse passive data collected ranging from mobile app usage to mobile sensing data. This type of mobile phone data can also be utilized on a larger scale (e.g., understanding people's behaviors and activities at the community level)~\cite{ghahramani2020urban}. Recent advances in mobile, wearable, and IoT technologies have enabled a wide array of novel health care consumer applications using big data analytics~\cite{imran2020big}.

\begin{figure*}[!t]
\centering
\includegraphics[width=0.8\textwidth]{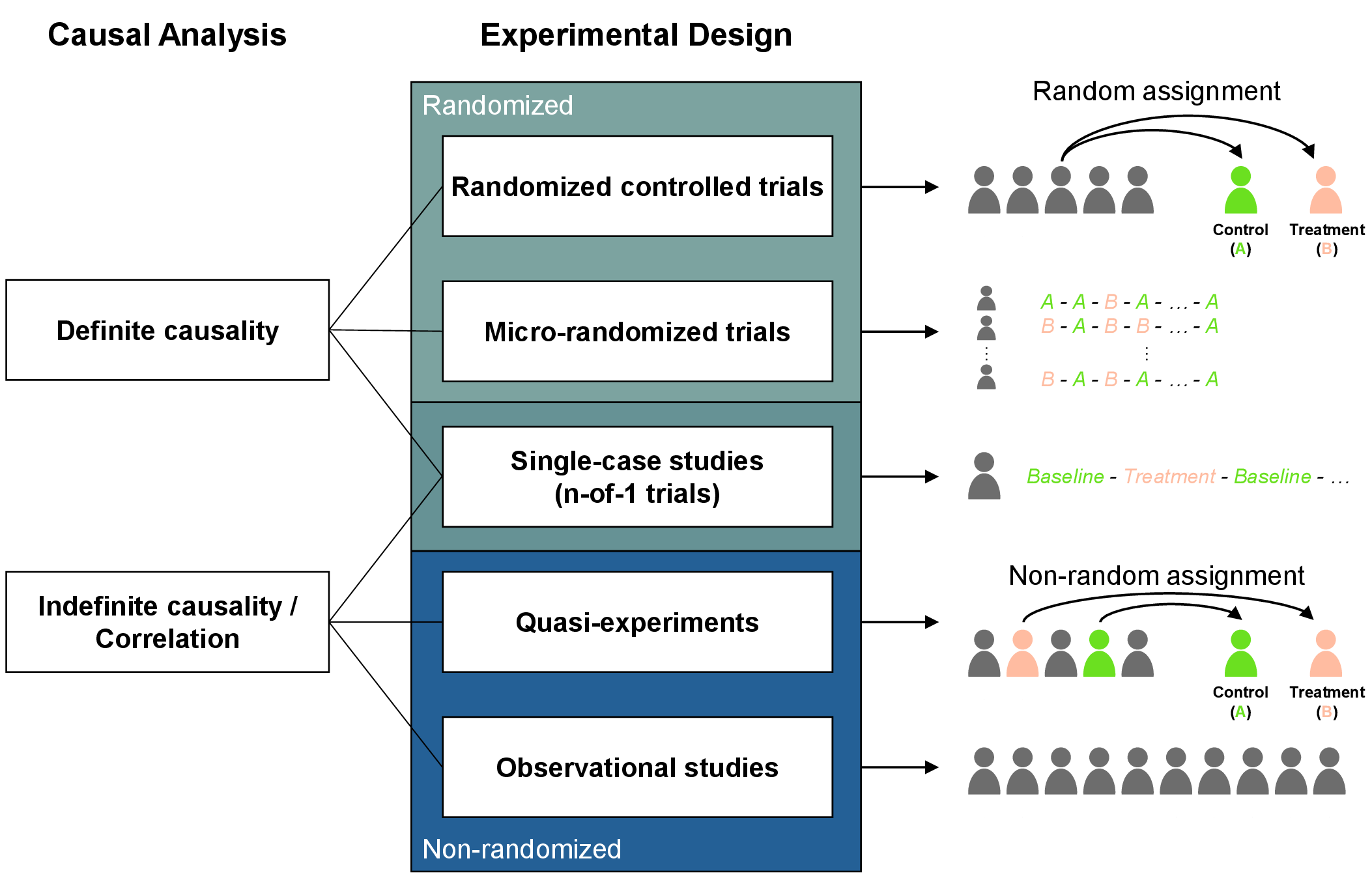}
\caption{Experimental design and data analysis framework of DTx}
\label{fig:dtxexperiment}
\end{figure*}

\begin{figure*}[!t]
\centering
\includegraphics[width=0.95\textwidth]{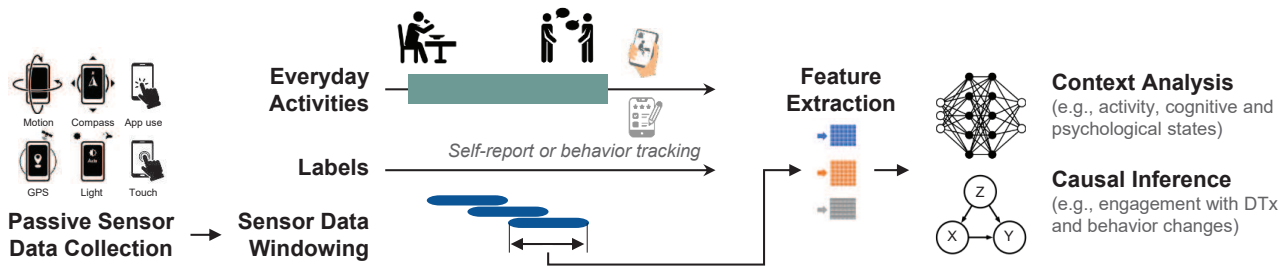}
\caption{Data analysis pipeline for DTx analytics}
\label{fig:mlpipeline}
\end{figure*}

\subsection{Experimental Design and Data Analysis Methods for DTx}

Similarly to the development procedure of conventional therapeutics, the development and application of DTx depend heavily on the choice of experimental design and subsequent data analysis (Fig. \ref{fig:dtxexperiment}). Various existing experimental designs suggested for eHealth or mHealth experiments can be applied to DTx to establish a causal relationship (i.e., evidence-based treatment). RCTs are the gold standard, whereby each experimental unit (or human subject) is randomly assigned to treatments (or conditions), and the outcomes are measured. The causal effect of the introduced treatment on the outcome variables is investigated by comparing the values of the outcome variables before and after the intervention while accounting for confounders. 

The experimental design of data-driven DTx analytics further involves continuous passive data collection. Beyond establishing a causal relationship between the treatment and outcome variables, passively observed sensor data provide additional opportunities for DTx optimization by extracting various context and intervention variables (Fig. \ref{fig:mlpipeline}). Passive sensor data analyses can be used to understand intervention contexts, improve delivery mechanisms, evaluate intervention components, and personalize treatments (see Section \ref{subsec:dtxdesign}). In reality, RCTs are very expensive as they require hundreds (or even thousands) of participants, and are even less practical when DTx software has already been deployed. In the following subsection, both randomized and non-randomized experimental designs are reviewed along with the associated context analysis and causal effect inference methods applicable to data-driven DTx analytics.

\subsubsection{Experimental Design for Data-Driven DTx Analytics} 
As with conventional therapeutics development, RCTs are the most appropriate experimental designs for DTx. In RCTs, the subjects are randomly assigned to either the experimental or control group. The treatment outcomes of the experimental group are compared with those of the control group to assess treatment effectiveness. In the RCT for DTx, the experimental group is given a digital or mobile platform, such as a smartphone application, and receives the desired intervention through the platform. In contrast, the control group can be defined in multiple ways: subjects in the control group may be given either conventional treatments, applications without any intended treatment effects, or no intervention at all~\cite{Firth2017}. Comprehensive reviews on RCTs for DTx regarding diseases such as depression, anxiety, and type 2 diabetes have been published~\cite{Firth2017, Firth2017a, Cui2016}.

Another type of randomized design applicable to DTx is the micro-randomized trial (MRT)~\cite{Klasnja2015, Liao}, which is a design proposed specifically for JITAI mHealth experiments~\cite{Nahum-Shani}. In MRTs, the intervention a subject receives is randomized at every decision point, that is, a point in time during the experiment where an intervention may be effective. As a result, the interventions for each subject are sequentially randomized hundreds to thousands of times during the experiment, making the experiment resemble a sequential factorial design. Repeated randomizations for every subject point to two main advantages of using MRTs in DTx experiments. First, the time-varying causal effects of treatments can be assessed for the outcomes of interest. Second, both between-subject and within-subject comparisons are allowed in MRTs, making the experiments efficient. Randomizations may be stratified so that the number of interventions a subject receives is sufficient for all conditions (e.g., the stress level of subjects)~\cite{Dempsey2015}.

In cases where an RCT is not applicable, designs such as quasi-experimental studies and observational studies, in which the assignment to the treatment or control group is not randomized, may be conducted for DTx. Quasi-experimental studies are experimental studies in which interventions are manipulated by the researchers without random assignments or full control over all extraneous confounding variables. Observational studies are non-experimental studies in which interventions are not controlled by researchers. Although these studies cannot draw any definite conclusions about the causal treatment effect because of the presence of confounders, they can be used to initially investigate the introduced interventions. For example, a mobile application to aid the treatment of diabetes may be provided to patients who would use the app freely without any explicit intervention by the researchers during the field trial~\cite{Littlewood2015}. Quasi-experimental studies such as pretest-posttest designs~\cite{Park2012, Nundy2014} or matching designs~\cite{Tsapeli2015}, \cite{Mehrotra2017}, have also been conducted to investigate the efficacy of mobile applications.

Single-case (also known as n-of-1) designs are a family of experimental designs applicable to DTx that contain aspects of both randomized and non-randomized trials. Single-case trials are experiments in which each subject acts as its own control, usually because the available sample size is very small. Similarly to MRTs, single-case trials capture the temporal dynamics within the study, as is often required by DTx studies. In general, single-case trials consist of two phases: the baseline and treatment. First, data are collected during the baseline period with no intervention, which serves as the control for the subject. Afterward, the intervention of interest is given to the subjects, and data for the same variables are collected. The causal effect of the intervention is estimated by comparing the treatment and baseline periods. In many cases, the baseline and treatment periods are repeatedly alternated (possibly with washout periods) for the ``replication'' of experiments or a clearer distinction of the causal effect of the treatment, by controlling for the confounding variables. In addition, different types of interventions or gradual changes in the intervention may be introduced after each baseline period, if necessary. When the baseline and treatment periods are randomly allocated within subjects, the trial is called an n-of-1 RCT. A more detailed review of single-case studies can be found in \cite{Dallery2013, McDonald2017}.

\subsubsection{Context Analysis Methods for DTx Analytics} 
The goal of context analysis is to use observational data to support exploratory tasks of investigating intervention contexts and behavioral routines or patterns, which can be captured using a set of observable characteristics of an individual through mobile and wearable devices such as activity trackers and smartphone loggers~\cite{jain2015digital,harari2016using}. The data sources range from passive sensor data (e.g., self-trackers and smartphone logging) and social media use to active self-reporting (e.g., mood and stress).

Mobile and wearable devices allow the performing of continuous and unobtrusive measurements of user contexts and help to infer users' health and behaviors~\cite{insel2017digital}. Prior studies have proposed several methods for analyzing users' intervention contexts and behavioral patterns. Here, ``context'' typically means ``any information that can be used to characterize the situation of a person''~\cite{dey2001understanding}. A context refers to a situation and environment in which a device or user is situated by a set of relevant features, such as human factors (e.g., user, social setting, and tasks) and physical factors (e.g., location and infrastructure)~\cite{schmidt1999there}. Harari et al.~\cite{harari2016using} proposed a three-dimensional context model, which included social interaction, daily activities, and mobility patterns. Behavioral patterns in each dimension are further defined based on the data processing of user contexts (e.g., the duration of social interaction). Similarly, Mohr et al.~\cite{mohr2017personal} proposed a hierarchical contextual feature model in which low-level sensor data were transformed into low-level features that constituted high-level behavioral patterns. Most low-level features, such as location and activity types, are human-interpretable. High-level behavioral patterns may include intervention contexts or behavioral patterns related to mobility patterns (e.g., number of significant places visited) or interaction patterns (e.g., daily phone use frequency).

For context sensing, motion sensors can be used to detect various types of physical activity, such as movement~\cite{van2013toward}, sleeping~\cite{abdullah2014towards}, eating~\cite{thomaz2015practical}, and agitation~\cite{alam2017motion}. For example, a user's sedentary state can be easily calculated by comparing the arithmetic difference between accelerometer samples. Eating gestures can be recognized by applying machine learning (e.g., random forest) to wrist motion data measured using smartwatches~\cite{thomaz2015practical}. GPS location traces can be processed to detect significant places (e.g., home and work) using clustering techniques~\cite{canzian2015trajectories}. In addition to physical activity tracking via sensors, prior studies used passive sensor data to infer cognitive and psychological states, such as depression~\cite{canzian2015trajectories} and social anxiety~\cite{huang2017discovery}. To identify the depressive symptoms of an individual, for example, mobility features (e.g., significant places visited) were extracted from smartphone GPS traces and personalized machine learning models with a support vector machine (SVM) classifier were tested~\cite{canzian2015trajectories}.
One way of optimizing intervention contexts is to provide timely delivery of intervention with mobile, wearable, and IoT technologies, owing to obtrusive modality usage (e.g., visual, vibration, and sound notifications). However, interruptive messages result in productivity loss, increased stress, and time pressure~\cite{Mark2005NoTask}, implying that less opportune delivery of intervention may lead to a low level of intervention adherence~\cite{Adamczyk2004IfNow}. A user's behavioral routines can be leveraged to find opportune moments (e.g., activity transition times)~\cite{Ho:2005:UCC:1054972.1055100}. A user's contextual model based on temporal and location contexts can be further used to define the complex rules for delivery timing.

\subsubsection{Causal Inference Methods for Data-Driven DTx Analytics} 
Data-driven DTx analytics focuses on estimating the treatment effect under the potential outcomes (Neyman-Rubin causal) framework~\cite{Holland1986, Donald2005}, whereby the potential outcomes (both factual and counterfactual) or a contrast (e.g., difference and ratio) are estimated either nonparametrically or by using models such as generalized estimating equation (GEE) and multilevel model (MLM). These models assume certain parametric regression form, in which the fitted model explains the causal effects of the treatment variables. For example, in MRTs, the average treatment effect under the potential outcome framework can be estimated through standard GEE or MLM regression, where the estimated coefficient of the treatment variable represents the mean difference in proximal outcomes between when the intervention is given and when it is not \cite{Klasnja2015}. Furthermore, various estimation methods tailored for MRTs, such as the centered and weighted least-squares method \cite{Boruvka2018}, have also been proposed as a direct application of the models, which may lead to unstable estimation due to time-varying or endogenous covariates \cite{Dempsey2015, Boruvka2018, Qian2019, Qian2019a}. Similarly, the GEE and MLM can be used to estimate the treatment effect in single-case studies with appropriate modifications \cite{VandenNoortgate2007, Daza2018}.

In the case of quasi-experimental or observational studies, it is difficult to identify definite causality because confounders are not controlled. Despite this limitation, it is common to compare the treatment and control groups to obtain a sense of the efficacy of an intervention~\cite{Park2012, Nundy2014}. For example, as a pretest-posttest regarding the effect of a digital intervention on cardiovascular risk factors in postmenopausal women with obesity, the outcome variable (e.g., blood pressure) was measured before and after the experiment for both the treatment and control groups, and the difference in outcomes between the groups was compared~\cite{Park2012}. A statistically significant difference between the groups suggests that the interventions are effective, although conclusions regarding the causal effect of the intervention cannot be made. Correlational analyses can also be conducted for the same purpose.

One possible approach for estimating causal effects using observational data is to adjust for the effects of confounders using methods such as stratification, matching, standardization, and weighting~\cite{Hernan2020}. Adjustment methods can be applied either nonparametrically by calculating the necessary estimates from the data or parametrically by fitting some models to obtain the estimates (e.g., parametric G-formula and marginal structural models). It is common to use propensity scores~\cite{Rosenbaum1983} (or other types of balancing scores) to estimate the probability that the subject is placed in the treatment group, given covariate values, instead of directly using the observed variables. These methods attempt to create a pseudo-population or matched population from an observational dataset such that the treatment and control groups display a similar covariate distribution. For example, in a behavioral study using smartphone sensor data, the causal effects of various lifestyle factors (mobile phone use) on stress levels (emotional state) were analyzed by creating a matched population from context variables extracted from the sensor data~\cite{Tsapeli2015, Mehrotra2017}. A practical tutorial on matching methods can be found in this reference \cite{Sizemore2019}.

Traditional adjustment methods are not generalizable to time-varying treatments and confounders, thereby resulting in biased estimates. To handle more complex longitudinal observational data with time-varying covariates, a more general class of models called G-methods for time-varying treatments was developed by Robins et al.~\cite{Robins2009}. These models, under appropriate assumptions, provide consistent estimates on contrasts of average treatment effects in scenarios with time-varying treatments and confounders by taking into consideration the ``history'' (the value of treatment and confounding variables up to the time point of interest) of subjects.

This review focuses on the potential outcomes framework. In contrast, the causal inference developed by Pearl~\cite{Pearl2009}, based on causal graphs and structural causal models, can also be implemented for similar purposes. In Pearl's framework, each node represents a variable, and the directed edges represent direct causation from one variable to another. The causal relationships encoded in \emph{causal graphs} can be translated into structural models and various causal inference tasks, such as causal structure discovery, which retrieves the underlying causal graph from observational data, and then treatment effect estimation from observational data can be conducted.

Another notion of causality often encountered in the literature is \emph{predictive causality}. Granger causality~\cite{Granger1969}, one of the most common concepts of predictive causality, measures whether a time series is predictive of another by testing whether the lagged values of one variable have predictive power for the upcoming values. In the case of dynamic systems where variables exhibit deterministic relationships, convergent cross mapping~\cite{sugihara2012detecting} can be used to discover the causality of two time series by leveraging time-delayed embeddings. These methods have been used to explore predictive causality in mobile data~\cite{twenge2018decreases, sarsenbayeva2020does}. Readers can find a comparative analysis of these approaches in~\cite{Yuan2020}.

\subsection{Data Visualization for DTx} 
Data visualization aims to visually represent datasets to help people perform exploratory tasks more effectively~\cite{Munzner2014Visualization}. Existing data visualization tools often support various user interactions such as changing visual encodings and navigating datasets (e.g., zooming, slicing, and projecting), thereby facilitating data-driven reasoning or decision-making, which is referred to as visual analytics~\cite{osti_912515}. 

Data-driven DTx analytics is based on personal big data collected from mobile and wearable devices. Passively collected DTx datasets are characterized by a high volume and velocity data-stream of sensor and interaction data with different spatiotemporal granularities (e.g., temporal: event time or day/week/month windowing, spatial: GPS points, places, or regions), semantic hierarchies (e.g., semantic grouping of sensors), and data types (quantitative, categorical, and unstructured). Visual analytics tools should support visual data exploration (e.g., charting datasets) and automated data analysis (e.g., data mining and machine learning)~\cite{osti_912515}. Visual data exploration, on the other hand, helps users check data quality and test research hypotheses visually. In addition, insightful information can be extracted by interacting with the data (e.g., examining and comparing different datasets and, zooming in on different parts). These insights, in turn, help build better models for data analyses and visualization explorations. In data-driven DTx analytics, this type of value-sensitive visualization is critical. Clinicians or developers may want to inspect user contexts, DTx usage, and behaviors not only to understand user engagement and behavioral adherence but also to acquire design insights for DTx improvements. 

When visualizing time-oriented data, the first step is to characterize the key characteristics of time-series data, such as temporal primitives (i.e., time points vs. intervals) and the structure of time (e.g., linear vs. cyclic as in seasons of the year)~\cite{Aigner207Visualizing}. As reviewed by Brehmer et al.~\cite{Brehmer2017Timelines}, effective timeline visualization requires careful exploration of the design space such as visual representation (e.g., linear, radial, grid, or spiral), scale of a timeline (e.g., sequential, chronological, relative, or logarithmic), and layout of views (e.g., single vs. multiple timelines, and segmented timelines). In addition, analytic reasoning for time-series data can be facilitated by supporting diverse user-interaction techniques (e.g., zooming, faceting, focus, and context)~\cite{Walker2016}. Thus, data visualization support for DTx must carefully consider data characteristics of mobile and wearable data and analytic reasoning tasks. 

\begin{figure*}[!t]
\centering
\includegraphics[width=\textwidth]{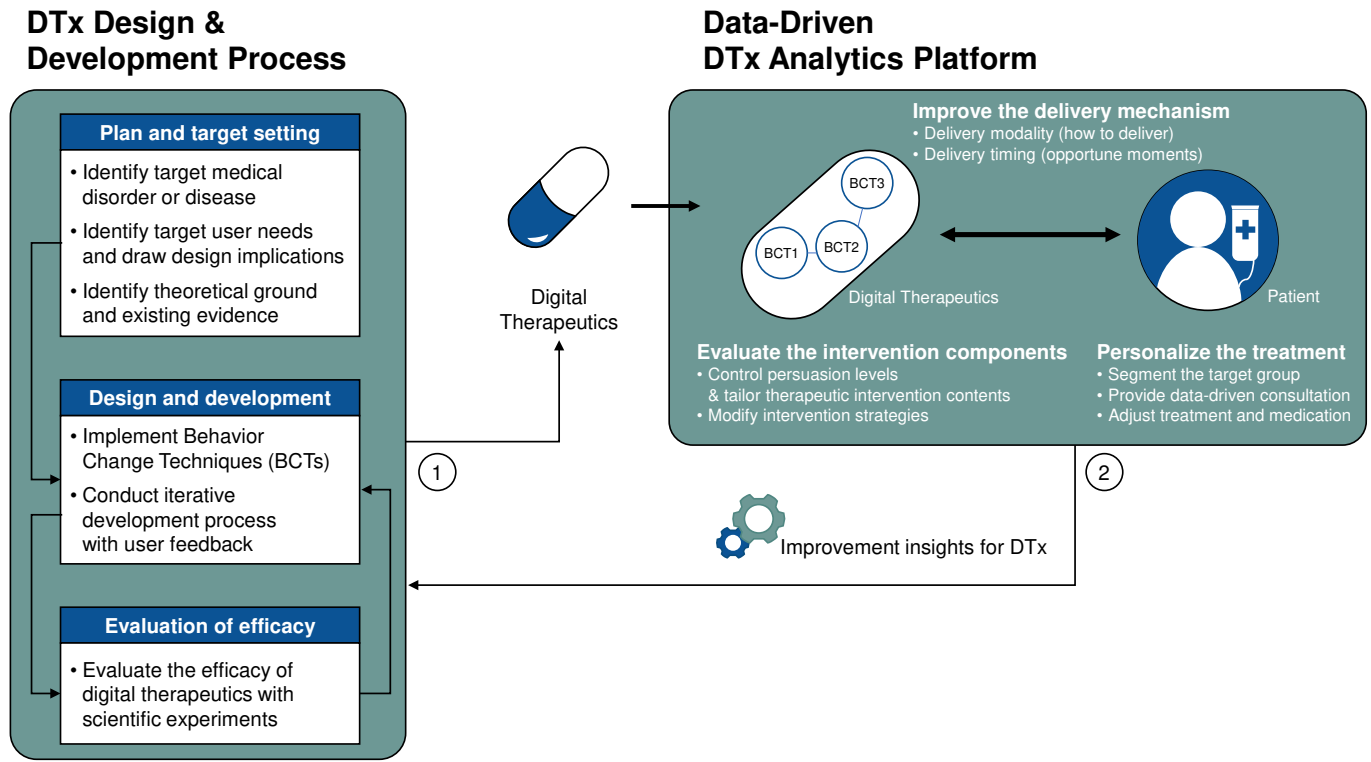}
\caption{DTx Design and Development Process Overview: (1) DTx design and development process with a) Plan and target setting, b) Design and development, c) Evaluation of efficacy, and (2) Data-driven DTx analytics for acquiring insights for improvement in three ways---a) Improve the delivery mechanism, b) Evaluate the intervention components, and c) Personalize the treatment. These insights can be re-fed back into the development process to iteratively improve DTx.}
\label{fig:dtxdesign}
\end{figure*}

Visualizing a large volume of personal big data is challenging; thus, it is important to support visual summaries of multidimensional time-series data. A simple visualization of the data streams collected from mobile and wearable devices would not be scalable; for example, a minute-level ``behaviorgram'' of smartphone data~\cite{Chen2019Cognitive} could be quickly cluttered as the number of data streams increases. For effective visualization, DataMD~\cite{DataMD2017} provides a visual dashboard for clinicians with visual summaries and trend lines of user activities for data-driven consultation. Interactive visualization allows users to compare multiple relevant timelines to help explore rich contextual data (e.g., identifying stressors by examining stress levels across different contexts over time~\cite{Sharmin2015Stress}). Advanced machine learning techniques can be used to find key relevant features such as behavioral markers related to cognitive decline~\cite{Chen2019Cognitive}. Event sequence mining can be used to uncover temporal patterns such as the event patterns before the occurrence of smoking~\cite{Chronodes2018}. 
 
\subsection{DTx Design and Development Process} 
\label{subsec:dtxdesign}

Digital therapeutics can be developed and improved during the product lifecycle. In this section, a review of DTx design and development processes is provided to illustrate how data-driven DTx analytics can help improve DTx services (Fig. \ref{fig:dtxdesign}).

\subsubsection{DTx Design and Development Process Review} 
The concept of DTx is highly related to that of digital health and persuasive technology in that software is used to persuade users to change their behaviors. Thus, a literature review is provided regarding the design and development process: (1) plan and target setting, (2) design and development, and (3) evaluation of efficacy.

\emph{Plan and target setting}: The first step is to identify the target medical disorder or disease. Fogg mentioned ``a simple target behavior to change'' as the first consideration for designing a successful persuasive technology~\cite{fogg2009creating}. Since the design process of DTx is mainly rooted in pervasive technology, a target disease or behavior should be selected first. The behavioral changes mentioned here could be the detailed goals of a larger objective (i.e., dealing with medical disorders). For example, patients with prediabetes can set detailed goals for exercise or diet management to prevent diabetes. After setting the target disease, researchers should identify the needs of the target user and draw design implications. It is crucial to understand the patients' point of view on digital health product use to implement what they need for the behavior change through ``patient-centered design'' methods, which involve patients in the process of development~\cite{birnbaum2015patient}. In addition, it is necessary to identify the theoretical basis and existing evidence underlying the design of the intervention, which provides the rationale for selecting specific target diseases and behaviors, target users, and behavioral change techniques~\cite{webb2010using}. Evidence from existing research can be collected using a systematic review of RCTs, or by testing relevant proven methods before development~\cite{mrc2019developing}. 

\emph{Design and development}: The second step begins with the implementation of behavior change techniques. The behavior change technique (BCT) refers to ``an observable, replicable, and irreducible component of an intervention designed to alter or redirect causal processes that regulate behavior.~\cite{michie2013behavior}'' For example, for the purpose of physical activities, various BCTs, such as providing guidance for a specific behavior, providing feedback on the user's performance, and setting goals of behavior and outcome are widely used~\cite{conroy2014behavior}. During this process, the number and type of BCTs that can be applied to DTx should be determined based on the identified target diseases, users, and theoretical/evidence bases. BCTs are then instantiated with certain intervention components to allow patients to interact directly with them. During this process, the medium to use, timing, and frequency of delivery of the intervention should also be considered~\cite{mohr2014behavioral}. As an iterative development process, pilot tests should be conducted on potential users (in this case, patients with the target disease) to examine the acceptability and feasibility of DTx intervention, user engagement, and adherence to the instructions.

\emph{Evaluation of efficacy}: The third step is to evaluate the efficacy of the DTx intervention. The assessment of the efficacy can measure different endpoints: (1) how much the patient behavior changed (i.e., proximal outcome) and (2) the degree of disease prevention or treatment (i.e., distal outcome). Randomized controlled (or clinical) trials are favored for assessing the relative effectiveness of DTx between treatment and control groups to reduce potential biases (e.g., selection bias). According to a report, RCTs and meta-analysis studies evaluating the clinical efficacy of digital health apps are gradually increasing~\cite{aitken2017growing}. 

\subsubsection{Data-Driven DTx Analytics for Optimizing DTx Delivery}
Data-driven DTx analytics can improve the efficacy of existing DTx by uncovering insights to improve therapeutics and apply them to the iterative development cycle based on the patient data. Since existing clinical trials have evaluated the efficacy of digital health interventions mainly with the difference in endpoints, it becomes a challenge to understand how certain elements of DTx induce differences and which factors beyond DTx affect patients' health concurrently~\cite{world2019guideline}. Beyond efficacy evaluation, data-driven DTx analytics can play an important role in further improving DTx by (1) improving the delivery mechanism, (2) evaluating the intervention components, and (3) personalizing treatments.

\emph{Improve delivery mechanisms}: For the delivery of health interventions, it is necessary to determine the channels of interaction between DTx and the patient. The appropriate channels of sensory input and output, or modalities, may vary depending on user preference (i.e., the preferred device or modality) or context (i.e., whether the intervention is palpable or interruption by sound notification is allowed)~\cite{lee2019intelligent}. The patient's context and the actual response for each modality can be analyzed in real time to suggest an effective method that most likely changes the patient's behavior. Modality selection can be automatic, considering the trade-off between the most effective modality and its availability. (e.g., converting sound notification into mute and LED light during a meeting). Depending on the context, the patient may or may not be aware of the DTx instructions and hence may not adhere to them. Therefore, it is crucial to detect ``opportune moments'' to interrupt patients to deliver timely health interventions to maximize efficacy. These moments can be determined by context-sensing methods specific to the platform or feature extraction by machine learning~\cite{lee2019intelligent}. During this process, sensor data may be utilized from smartphones (e.g., ambient light, sound, or acceleration of the device) for interaction (e.g., touching or sliding the screen) to understand the context of the user, as suggested in~\cite{chen2019behavioral}. It is also possible to divide the delivery process of the intervention into several stages to identify the point at which the process fails~\cite{choi2019multi}. As suggested in this study, whether the patient (1) perceives the intervention, (2) is available to respond, (3) adheres to the intervention, and (4) performs the target behavior can be affected by several contexts. Thus, contextual data are collected and utilized to build a model that can improve delivery mechanisms.

\emph{Evaluate the intervention components}: During the development process of DTx, BCTs are determined based on a theoretical basis or existing proven research. However, such BCTs may not result in the same efficacy because of differences in the target medical disorder, target user traits, or the context of therapeutic use. Data-driven DTx analytics can be used to evaluate the extent to which each BCT component affects the target behavior (proximal outcome). A component-wise evaluation is less practical because of its complexity and cost. Instead, the observed sensor data can be used to determine this type of component-wise efficacy via a pseudo-experiment design. This helps determine the relative importance of components and guides designers and clinicians on how to improve the components or to better evaluate BCTs (experimental design). In addition, it may be possible to identify inter-relationships and causal relationships of the components, similarly to a model describing correlations and interactions among health-related attitudes~\cite{orji2012towards}. This evaluation can help develop a new strategy, such as emphasizing certain BCT components by similar patient groups or deploying motivators to enhance their use. Compliance with persuasion for DTx may vary depending on the patient's personality traits, such as extraversion, agreeableness, conscientiousness, neuroticism, and openness to experience~\cite{alkics2015impact}. The data collected can be used to create a ``persuasion profile'' to predict how many different types of persuasion principles (e.g., authority, consensus, commitment, or scarcity) would affect patients with different personalities~\cite{kaptein2015personalizing}. Thus, the persuasion level can be controlled by changing the intensity of the wording, the frequency of sending persuasive messages, and personalizing persuasion principles to each individual. In addition, the patterns of interaction can differ with intervention content because the patient may have difficulty in understanding the content, lose interest in the content, or not be motivated to learn. Data-driven analytics then utilizes interaction patterns to design intervention coursework or recommend content that may engage the patient's interest. Recently, Yue et al.~\cite{yue2021overview} reviewed widely used techniques for content recommendation, such as content-based (i.e., content with similar features), collaborative filtering-based (i.e., modeling user behavior), and hybrid methods. These approaches can also be adapted to the intervention content recommendations in DTx.

\emph{Personalize the treatment}: DTx can be further improved or ``personalized'' through an understanding of the target group via contextual analytics. Patients can be divided into subgroups based on their engagement in or adherence to therapeutics. For example, Alshurafa et al.~\cite{Alshurafa18IsMore} classified clusters of patients who showed similar health behavior changes in four categories (i.e., ``healthy and steady,'' ``unhealthy and steady,'' ``decliners,'' and ``improvers''). This approach may help predict changes in health activities by considering the characteristics of patient groups to formulate appropriate strategies. In addition, more personalized treatments can be achieved by adding personal traits. For instance, the outcome of behavior change can be analyzed in relation to certain variables, such as demographics, health status, and psychosocial features~\cite{boslaugh2005comparing}. This process enables the analysis of more specific target segments and even covers niche patient groups using upgraded DTx. The analytics platform can also help health coaches understand patient engagement and adherence to DTx based on collected data (e.g., the patients' context, therapeutic usage pattern, and proximal and distal outcomes). Moreover, the platform may suggest the advice needed for the patient by visualizing activity data. With this information, the coach can provide data-driven, personalized discussions to the patient and improve the efficacy of therapeutics. DTx can be used independently or along with other traditional treatments. If the patient takes medicine as part of DTx (e.g., Pear Therapeutics' substance use disorder intervention), the analytics can be used to increase the effectiveness of the medication in use and result in better outcomes. For example, in the case of depression, DTx allows the patient to record the degree of depression, quality of sleep, and activities performed. Based on these data, a clinician can control the type, combination, or dose of medication when the patient visits the hospital.
 
\section{Research Directions for Data-Driven DTx Analytics} 

\subsection{Data Collection and Data-Driven Delivery Optimization} 

\emph{Data quality management}: While conducting a large-scale experiment in the wild, a general-purpose data platform guarantees that all necessary data from users are collected with data quality support (e.g., format, completeness, and freshness~\cite{askham2013six}). When data quality is compromised, the analytics platform can notify key stakeholders (e.g., data providers, managers, or DTx developers) for further intervention (e.g., checking mobile apps or fixing software errors). Moreover, the analytics platform may need to detect a user's abnormal behavior for quality management (i.e., injecting fake data to maximize rewards or violation of experimental guidelines). These abnormalities can be easily captured by outlier detection with suitable feature tracking (e.g., local outlier detection~\cite{Breunig:00} and machine learning~\cite{malhotra:2015}). Existing CTMSs or data collection platforms can consider implementing additional features to enhance data quality management. 

\emph{Intervention delivery optimization}: Another research direction related to receptivity improvement of analytics platforms is opportune moment detection and modality selection; these analyze tailoring variables (i.e., user's behavioral and contextual data) and intervention components to find the most effective time that maximizes user engagement. This includes deferring the intervention until the predicted deadline~\cite{Mehrotra:2019}. Furthermore, the analytics platform requires selecting an appropriate modality in a multi-device environment (e.g., laptops, smartphones, smartwatches, and IoT appliances) to consider user preferences, device accessibility, and properties of the intervention~\cite{weber2016situ}. Finally, finding an opportune moment and modality selection should be cross-optimized to improve receptivity to the DTx software, which is an open and interesting research area.

\subsection{Causal Inference for Data-Driven DTx Analytics}

In DTx scenarios, observational studies are conducted along with randomized studies because of the cost or difficulty of conducting complex experiments. For example, the number of conditions increases exponentially with the number of factors (e.g., intervention components). Recent advances in artificial intelligence and deep learning have led to high-performing causal inference models for observational data, which handle the lack of randomized controlled environments through model architecture~\cite{Hill2011, Shalit2017, Alaa2017, Wager2018, Yoon2018, Shi2019}. In these studies, individual treatment effect estimation was performed through counterfactual inference, in which machine learning models, including deep neural networks~\cite{Shalit2017, Alaa2017, Yoon2018, Shi2019} and tree-based models \cite{Hill2011, Wager2018}, were used to estimate all potential outcomes and thus the causal effect. Counterfactual inference models that use sequential models such as recurrent neural networks have also been proposed to handle the bias from time-varying confounders~\cite{Lim, Bica2020, liu2020estimating}. Similar counterfactual inference frameworks can be applied to data-driven DTx analytics in which the effect of treatments delivered through digital or mobile platforms is estimated through deep networks based on passively collected DTx datasets. Various definitions of treatments can be explored and implemented accordingly, depending on the causal analysis of interest, through effect estimation models. For example, treatment variables can be defined based on the engagement levels of the DTx software, or various sets of actions can be observed via passively collected data. Regardless of how the treatment variables are defined, an appropriate model architecture that incorporates all aspects of passively collected DTx data is required to estimate all potential outcomes for accurate causal analysis of DTx. 

\subsection{Data Visualization for Data-Driven DTx Analytics}

One of the critical challenges in data visualization is the volume of personal stream big data. During a clinical trial, a stream of sensor and interaction data can be collected from hundreds or thousands of people over a few months. As discussed in Section \ref{sec:ch:design}, it is also possible to collect data after deployments for continuous improvements (from those who opt-in for data collection). Prior studies explored various interactive visualization techniques to efficiently navigate through large-scale disease or patient datasets~\cite{CareGiver2005, CareFlow2013}; for example, CareFlow helps visualize the outcome data of 50,000 patients in a tree-like timeline and provides a high-level overview of similar groups of patients~\cite{CareFlow2013}. Large-scale datasets are processed to extract hundreds or thousands of features for contextual and causal analytics. Visualization of such high-dimensional data is challenging. As in existing machine learning studies, dimensionality reduction techniques such as principal component analysis or t-distributed stochastic neighbor embedding can be used for dataset exploration and annotation. Leveraging domain knowledge as well as prior data-driven DTx analytics experience can also help focus on specific sensor and interaction data types. As illustrated earlier, DTx design and development are mainly led by prior domain knowledge on theoretical grounds and existing evidence~\cite{webb2010using}. According to a recent survey on depression, several key behavioral features include location, physical activity, and sleep~\cite{Rohani2018Correlations}. Data-driven DTx analytics, including correlation analyses and causal inference, allows for further reduction of the feature space. As a result, clinicians can manually examine the reduced feature space to effectively explore hypotheses on contextual and intervention factors affecting engagement and behavioral adherence. It is envisioned that artificial intelligence-inspired ranking mechanisms will automatically generate a ranked list of hypotheses to be examined.

\subsection{DTx Design and Development Process}

\label{sec:ch:design}
The existing literature on DTx design mostly focuses on design principles, and there is a lack of practical guidelines on how software development and evaluation methods can be incorporated into the design and development process. As shown earlier, the US FDA introduced a pre-certification pilot program that certifies SaMD ``developers'' instead of SaMD ``products'' to manage the total product lifecycle (TPLC), which ``enables the evaluation and monitoring of a software product from pre-market development to post-market performance, along with the continued demonstration of the organization's excellence''~\cite{FDA-PreCert}. It is important to collect and analyze real-world datasets for DTx to continuously improve product safety and effectiveness and to deal with potential risks. 

The challenge is to establish user-centered agile software development (UCASD) practices and principles for DTx design, development, and evaluation, as recommended by Pach et al.~\cite{Pach2021} (e.g., minimizing the complexity of DTx design and jointly conducting app development and evaluation). Well-known UCASD principles can be applied to DTx. Software design and development occur in short iterations, with incremental improvements. Design, development, and evaluation occur in parallel interwoven tracks, and stakeholders (e.g., patients, clinicians, and developers) are actively involved in the entire process~\cite{Brhel2015Exploring, Lord2021}. In UCASD, evaluation can occur at various stages: initial (pre)design phases (e.g., plan/target setting and implementation of behavior change techniques), iterative development of DTx software, clinical evidence evaluation (for clinical trials), and in-the-wild deployments. Existing user-centered design techniques (e.g., small-scale user studies by interviewing participants) can consider ``data-driven DTx analytics.'' It is important to systematically examine how diverse datasets collected during different evaluation stages can be used by key stakeholders to acquire design insights for DTx improvements (e.g., data mining for usability issue identification~\cite{Gonzalez2008Enhancing}). For example, during the clinical evaluation stage, RCTs can be conducted using passive data collection for data-driven analytics. When considering the TPLC of DTx, data-driven analytics helps a company not only with continuous monitoring of product safety and risk factors but also in providing opportunities for iterative design improvements. Further case studies on DTx design and development processes using data-driven DTx analytics should be conducted to establish practical DTx design and development practices and principles.

The review and discussion in this work mostly focused on conventional mobile and wearable application designs for DTx. It is increasingly important to consider new computing platforms for BCT delivery, such as virtual reality and metaverse (e.g., Facebook's Horizon Workrooms, Microsoft Mesh, and NVIDIA Omniverse)~\cite{Gold2021VR} and conversational agents (e.g., Google Assistant)~\cite{Car2020}. Further studies must be conducted to understand user engagement and adherence patterns in novel platforms, such as tracking an avatar's behaviors in the metaverse for data-driven DTx analytics. In addition, the delivery of therapeutic interventions can be extended using physically embodied smart agents, such as robots. In terms of human-robot interaction, diverse responses are collected from users, such as visual/vocal nonverbal cues or subjective feelings during the interaction, which are then utilized to evaluate and improve interventions~\cite{shen2020understanding, liu2017facial}.

\subsection{Privacy and Ethics in DTx Analytics}

Given the DTx analytics uses a vast amount of personal data collected from mobile, wearable, and IoT devices (e.g., GPS locations, phone usage, and emotion samples), privacy and ethics must be carefully evaluated. Recent studies examined user motivations and concerns when collecting data on mobile and wearable devices~\cite{Rooksby2019DigitalPhenotyping, Lee2022IMWUT}. The two main reasons why users provide their data are financial gain and altruistic benefits (e.g., scientific advances)~\cite{Lee2022IMWUT}. According to Rooksby et al.~\cite{Rooksby2019DigitalPhenotyping}, the major concerns are related to negative user experiences of data collection; e.g., privacy concerns of sensitive data collection, negative thoughts and feelings about self-reporting, and detrimental effects on device performance. Although the majority of data contributors are not typically concerned with privacy, Lee et al.~\cite{Lee2022IMWUT} further showed that privacy concerns are related to feelings of surveillance, the identification of daily routines, and data breach. 

For ethical reasons, Huh-Yoo et al.~\cite{HuhYoo2021Pervasive} stressed the significance of communicating with the study participants as part of the informed consent process by elaborating on the known potential risks (as well as their additional concerns) and how to mitigate those risks. In DTx settings, it is critical to communicate the types of behavioral and sensor data that are gathered, whether any of it contains any personally identifiable information, and the potential risks associated with sharing sensor data. However, it was found that users frequently have incorrect mental models regarding the types of data collected from mobile and wearable devices, and how they are used to build AI models~\cite{Lee2022IMWUT}. Therefore, researchers must provide intuitive explanations about the types of data collected and allow users to access their own data by supporting interactive data visualization tools (e.g., minute-level behaviorgram of sensor data streams~\cite{Chen2019Developing}). In fact, DTx analytics could leverage the MyData vision that empowers users by offering \emph{direct} access and control of their own data~\cite{Alorwu2021MyData}. The MyData vision extends the EU's General Data Protection Regulation (GDPR) that intends to enhance users’ control and rights over personal data. Further study is required to develop novel user-friendly tools that allow users to directly access and control their own data as part of DTx data analytics platforms, similar to Privacy Bird~\cite{Cranor2006PrivacyBird}, a user-friendly web data privacy management tool for the Platform for Privacy Preferences (P3P).

Data collection requires informed consents from participants where researchers and clinicians must inform participants about the risks and benefits of study participation. Before collecting any data, one-time informed consents are normally obtained. Recent research has supported the usage of dynamic informed consents, which enable users to adaptively modify their consents in response to their contextual needs, such as turning off GPS while visiting hospitals~\cite{Lee2021DynamicConsent}. Future research should investigate the design space of dynamic consents in mobile and wearable environments, much like earlier studies that thoroughly studied the design space of privacy notice and choices~\cite{Schaub2015PrivacyNotice, Feng2021PrivacyChoices}. We anticipate that implementing dynamic consents will enhance the MyData vision in DTx contexts.

\section{Conclusion} 
Digital therapeutics as SaMD aims to cure diseases and improve health conditions, which is a major departure from existing wellness products. Health interventions are delivered through digital technologies (e.g., mobile content, chatbots, and push notifications); thus, it is critical to analyze and optimize the engagement and receptivity to DTx delivery systems. We proposed a data-driven DTx analytics framework that allows researchers and practitioners to collect mobile, wearable, and IoT data as part of field experiments and to perform contextual analytics and causal inference for DTx delivery optimization. We reviewed the core components of data-driven DTx analytics such as data collection and user management, data-driven modeling for DTx usage and behavioral adherence, experimental design and contextual/causal analytics, data visualization, and DTx design optimization. Finally, we discussed research directions for data-driven DTx analytics, such as causal inference, data visualization, delivery optimization, design process optimization, and the privacy and ethics of data-driven DTx. Over the last few decades, there have been significant advances in mobile sensing and machine learning, and observational data have provided new opportunities for DTx development and optimization. However, data-driven DTx analytics faces novel challenges such as privacy risks due to extensive data surveillance and development and operation risks due to the heterogeneity of mobile platforms and IoT devices (e.g., iOS vs. Android). Despite these challenges, we hope that our work serves as a foundation for this new research direction, and we call for further research on data-driven DTx analytics.

\section*{Acknowledgment}
This research was supported by Basic Science Research Program through the National Research Foundation of Korea (NRF) funded by the Korea government (MSIT) (2020R1A4A1018774)

\ifCLASSOPTIONcaptionsoff
  \newpage
\fi

\bibliographystyle{IEEEtran}
\bibliography{references}

\end{document}